\documentclass[final,3p,times]{elsarticle}
\usepackage{geometry}
\usepackage{lipsum}
\usepackage{graphicx} 
\usepackage{caption}
\usepackage{subcaption} 
\usepackage{amsfonts, amssymb, amsmath} 
\usepackage{float} 
\usepackage{placeins} 

\usepackage{threeparttable} 
\usepackage{array, booktabs, siunitx}
\usepackage{lineno}
\usepackage{comment}

\usepackage{ragged2e} 
\usepackage{hyperref} 
\usepackage[normalem]{ulem} 

\usepackage{makecell}
\usepackage{verbatim}

\newcommand{\beginsupplement}{%
    \setcounter{table}{0}%
    \renewcommand{\thetable}{S\arabic{table}}

    \setcounter{figure}{0}%
    \renewcommand{\thefigure}{S\arabic{figure}}

    \setcounter{equation}{0}%
    \renewcommand{\theequation}{SI Eq. S\arabic{equation}}

    \setcounter{section}{0}%
    \renewcommand{\thesection}{SI Text S\arabic{section}}
}

\begin{document}
\begin{frontmatter}



\title{Sectoral inter-dependencies drive the loss of structural balance in signed financial networks}


\author[first]{Kartik Dahake}
\author[second,third]{Abhijit Chakraborty\corref{cor1}}
\ead{abhijit@labs.iisertirupati.ac.in}

\cortext[cor1]{Corresponding author}
\affiliation[first]{
organization={Department of Physics, Indian Institutes of Science Education and Research Tirupati},
city={Tirupati},
postcode={517619},
state={Andhra Pradesh},
country={India}
}

\affiliation[second]{
organization={Department of Humanities and Social Sciences, Indian Institutes of Science Education and Research Tirupati},
city={Tirupati},
postcode={517619},
state={Andhra Pradesh},
country={India}
}

\affiliation[third]{
organization={RIKEN Interdisciplinary Theoretical and Mathematical Sciences Program},
city={Wako},
postcode={351-0198},
state={Saitama},
country={Japan}
}

\begin{abstract}
Signed graphs provide an effective architecture for portraying a system in which cooperation and conflict coexist. Emerging from the concept of balance in psychological sciences, they have found applications across several domains. Financial markets are one such intriguing example that can be modeled using signed networks, where assets exhibit correlations in their price movements. During periods of systemic risk, such a signed financial network shows a loss of balance, which has been consistently demonstrated. Here, we specifically explore how this structural imbalance is distributed across different scales within the financial network, therefore revealing its mesoscopic origin. Adopting the framework of structural balance theory, we utilize a measure of polarization based on triadic motifs to investigate the distribution of structural imbalance across varying sectoral scales. Specifically, we analyze the temporal evolution of global polarization and its sectoral constituents using longitudinal data derived from the S\&P 500 index, a widely followed stock market index that tracks the performance of the largest 500 publicly traded companies in the United States. By decomposing global polarization into its intra-sectoral and inter-sectoral constituents, we show that structural imbalance arises predominantly from interactions between sectors rather than within them during periods marked by systemic risk. To establish a baseline on which observed structural imbalance in the network is to be assessed, we employ randomization protocols to confirm that observed imbalance configurations are statistically significant and not artifacts of lower-order interactions. Furthermore, we derive a regression equation demonstrating that the variance in global polarization is well explained by macroeconomic variables, indicating that low levels of global polarization during economic crises are driven by compounding pressures from supply chain disruptions and inflation uncertainty. Collectively, these findings provide a quantitative framework for understanding how localized sectoral conflicts propagate across the financial network and ultimately contribute to large-scale structural instability during periods of economic crisis. 
\end{abstract}



\end{frontmatter}



\section{Introduction}
\label{introduction}
A quintessential example of complexity is the financial market, in which numerous interacting market participants collectively give rise to the global market dynamics that are manifested through changes in measurable quantities such as stock prices, market indices and volatility~\cite{mantegna1999introduction}. The interactions among these market participants fundamentally govern the state of the market through a large number of daily transactions. These interactions are influenced by several factors, including individual investor preferences, external events such as geopolitical developments and macroeconomic announcements~\cite{gabaix2006institutional}, and herding behavior~\cite{cont2000herd}. The structure of interaction among the entities in a complex financial system can be revealed by focusing on the spectral properties of the empirical correlation matrix. A fundamental challenge addressed by early econophysics studies~\cite{laloux1999noise, plerou1999universal} was determining whether an empirical correlation matrix contains statistically significant structural information beyond what is expected from a purely random matrix representing uncorrelated price fluctuations. By measuring deviations from the predictions of Random Matrix Theory (RMT), researchers can identify non-random properties, thereby revealing the genuine information about the underlying structure of interaction~\cite{plerou2002random, kim2005systematic, pan2007collective, shen2009cross}. While the interdependencies among individual components give rise to the collective market dynamics, they inherently render the system susceptible to systemic risk~\cite{battiston2012debtrank, chakraborty2024inequality}; the phenomenon where a failure of a single or a few entities triggers a cascading effect that leads to entire market failure. Literature in this context has observed an increase in market correlations during periods of market instability and a declining trend in the structural centrality of the financial sector over the decade leading up to the 2008 Global Financial Crisis. ~\cite{aste2010correlation}. Furthermore, the temporal evolution of the interaction structures between distinct business sectors has been observed to change dramatically during such financial crises~\cite{jiang2014structure}. 

\begin{figure}[!htbp]
    \centering
    \includegraphics[width=0.8\textwidth]{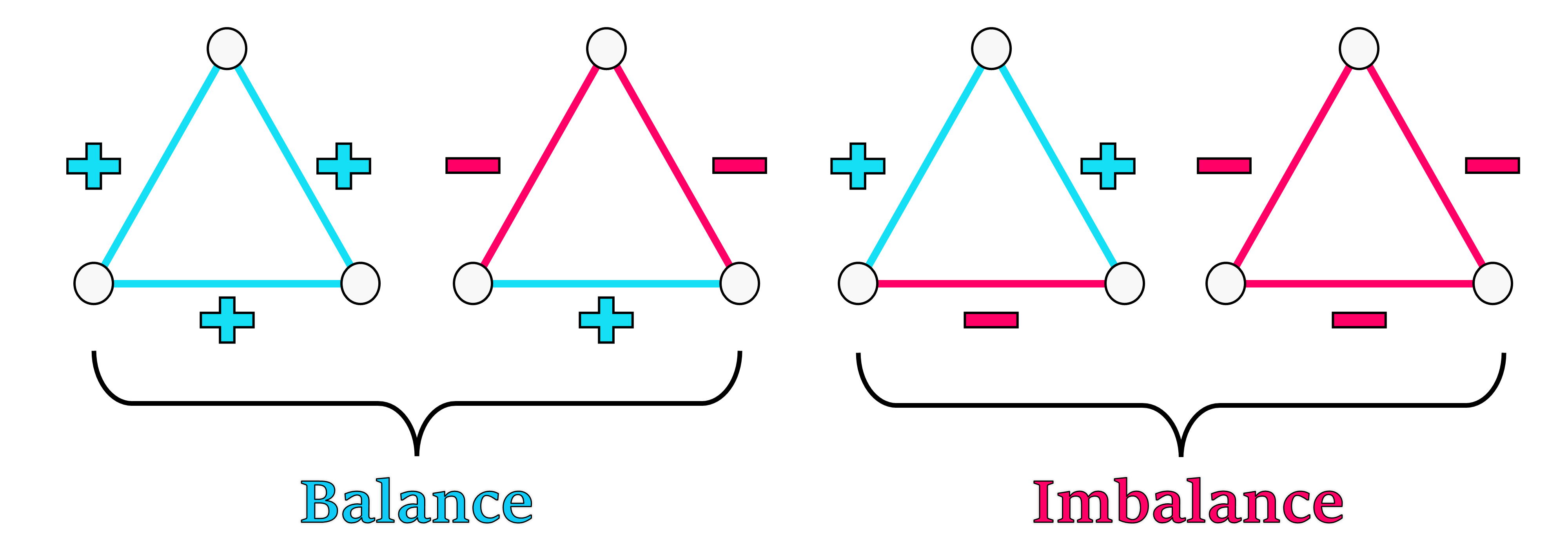}
    \caption{\textbf{Four possible signed triadic configurations in structural balance theory.} Blue edges denote positive correlations and magenta edges denote negative correlations between asset time series. Triads with an even number of negative edges, $(+++)$ and $(+--)$, are structurally balanced, whereas triads with an odd number of negative edges, $(++-)$ and $(---)$, are structurally unbalanced (frustrated).}
    \label{fig:all_triad_configs}
\end{figure}

Recent studies~\cite{kuyyamudi2019emergence, ferreira2021loss} have demonstrated that these systemic crisis events are closely linked to a loss of structural balance, a phenomenon signifying the emergence of frustration that leads to a major regime transition. The concept of structural balance was introduced in psychological science~\cite{heider1946attitudes}. Heider's balance theory considers higher-order interactions that extend beyond simple pairwise links. Later, Cartwright and Harary formalized the idea mathematically by using signed graphs. A signed graph is said to be structurally balanced if all of its cycles are positive~\cite{cartwright1956structural}. However, evaluating just the three-node interactions serves as an effective, first-order metric for global balance, since longer cycles contribute less to the overall network balance~\cite{marvel2009energy}. Because structural balance theory is fundamentally based on these triadic motifs, any two connected entities exhibit either a cooperating (positive) or conflicting (negative) relationship. A three-body interaction with each pairwise relation having two possible signs leads to four possible triadic configurations, as illustrated in \autoref{fig:all_triad_configs}. These include either all three positive relationships $(+++)$ or one positive and two negative relationships $(+--)$, collectively referred to as balanced triads in which an even number of negative relationships (0 or 2) exist. Conversely, a triad is unbalanced, or frustrated, if it contains an odd number of negative relationships: either two positive and one negative relationship $(++-)$ or all negative relationships $(---)$. In this regard, balance theory has been successfully adapted across various disciplines. M.~Zahedian \emph{et al.}~\cite{zahedian2022financial} employed the framework of structural balance theory and found that the financial network structure showed greater resilience to disorder during crisis periods than in non-crisis periods. Moving beyond pure topology, signed networks have also been modeled using Hamiltonian functions, providing a formulation of balance theory in terms of an energy landscape. For example, Marvel \emph{et al.}~\cite{marvel2009energy} expanded upon Cartwright and Harary's notion of global balance and showed that the energy landscape of signed networks is complex and filled with local minima, the so-called jammed states. A jammed state is a signed configuration wherein a change in the sign of a single edge increases the energy of the whole system. Furthermore, they derived bounds for these jammed states and proved that such states exhibit a modular structure. In another study employing the same statistical-mechanics framework of structural balance theory, Belaza \emph{et al.}~\cite{belaza2017statistical} proposed a three-term energy function that accounts for three-body interactions, node heterogeneity, and negative edges in the triads. Validating their model against both virtual and real-world datasets, they found a persistent hierarchy among the energy states of the four triad configuration. This Hamiltonian formulation of structural balance is equivalent to energy minimization of a spin glass~\cite{facchetti2011computing}. Spin glass materials are classic examples that exhibit a state of frustration~\cite{fischer1993spin}. The magnetic moments of atoms in the material are  randomly oriented and interact in a competing manner, preventing the system from settling into a minimum energy state. Just as physical systems attempt to resolve this frustration by aligning the magnetic domains, a complex signed network tends to reduce the structural tension by reorganizing into factions with internally consistent positive interactions and conflicting negative interactions between factions. This clustering mechanism gives rise to polarization, which has been extensively explored in social systems to study how varying the connectivity within these networks can trigger phase transitions in the system's overall polarization~\cite{minh2020effect, thurner2025more}.

Previous literature has established the mechanism leading to the loss of structural balance in financial networks, but the mesoscopic origin of this structural imbalance remains largely unexplored. Kuyyamudi \emph{et al.}~\cite{kuyyamudi2019emergence} demonstrated that the emergence of frustrated triads serves as an indicator of systemic risk on a global scale, providing a signature for distinguishing the transition in topology. Further, Ferreira \emph{et al.}~\cite{ferreira2021loss} showed that the observed balance-unbalance transition during the stock market crash is triggered by the formation of full-negative cliques composed of low capitalization stocks that migrate from the periphery to the center of the network. Although these studies identify the factors responsible for the structural imbalance, they do not address the systematic distribution of frustration across established economic sectors. We resolve this ambiguity by examining the mesoscopic organization to determine the exact origin of the global structural imbalance. 

This establishes the primary objective of our study: to perform a multiscale investigation to quantify how the loss of structural balance is distributed within and across sectors, which ultimately accumulates to give rise to the global polarization in the system. For this, we analyze the evolution of collective dynamics, especially during large-scale disruptions in one of the major stock market indices in the USA, viz., Standard and Poor's 500, over a longitudinal 14-year timeline that captures both relatively stable market conditions and the global disruption induced by the COVID-19 crisis. We undertake a temporal evolution study from the lens of structural balance theory to track the continuous evolution of market polarization across multiple sectoral scales to explicitly determine the origin of loss of structural balance in the financial network. To achieve this, we employ RMT to filter noise from cross-correlation matrices derived from asset returns, with particular emphasis on group correlation matrices to investigate sectoral inter-dependencies during calm and crisis periods. We implement the framework of structural balance theory, in which positive correlation indicates correlated movement between assets or sectors, and negative correlation indicates anti-correlated movement. Our approach encompasses a multiscale analysis that provides a comprehensive assessment of structural balance over time, thereby offering deeper insight into the sectoral dynamics underlying systemic risk.

\FloatBarrier
\section{Material and Methods}

\subsection{Data analyzed}
\label{subsec:data_collection_net_construction}

We conduct our analysis on the dataset consisting of the daily adjusted closing prices of the companies listed on the Standard and Poor’s 500, collected from \href{https://finance.yahoo.com/}{Yahoo Finance}~\cite{yahoofinance_covid_pipeline} using all available tickers corresponding to S\&P 500 constituents, with the dataset spanning from January 4, 2010 to December 30, 2024. Sectoral affiliation of S\&P 500 companies has been defined using the Global Industry Classification Standard (GICS)~\cite{msci2023gics} obtained at the time of data collection and provides the basis for the sectoral framework utilized throughout this study. Although MSCI and S\&P Dow Jones Indices annually review the GICS structure, we employ a static sector mapping, in which the sector assignment is fixed throughout each study period. This ensures a consistent basis for temporal comparison and allows observed changes in the network structure to be attributed to evolving market dynamics rather than to changes in sector classification. To ensure a consistent longitudinal sample, we identify a set of stocks that are consistently present across the 14-year timeline and have no more than 25 missing observations. This reduces the number of stocks to 426. The resulting longitudinal timeline is then analyzed by segmenting it into 56 overlapping windows. Each window spans 252 trading days and has a 63-day overlap. The epoch length of $T=252$ (one trading year) was deliberately chosen because shorter epochs provide a better approximation to stationarity, allowing the model to capture critical state transitions~\cite{pharasi2019complex}. To ensure data quality in each window, missing prices were forward-filled with the last available closing price, treating them as non-trading days. When missing values occurred at the beginning of a series with no prior prices, a backward-fill approach was applied. To further evaluate the robustness of the proposed framework, we also analyze an independent dataset comprising the daily adjusted closing prices of stocks listed in the S\&P 500 index over the period 2005-2015. This time frame encompasses both the periods of market stability and major economic crisis, notably the 2008 Global Financial Crisis (2008 GFC). A comprehensive presentation of the robustness analysis is provided in the Supplementary Information (SI) Text S1.

In the final part of the empirical analysis, we employ a multiple linear regression (MLR) model to examine how the proposed polarization measure is associated with broader economic conditions. To this end, we consider two macroeconomic variables: The Global Supply Chain Pressure index (GSCPI), which quantifies the state of the global supply chain by combining a range of transportation and manufacturing-related indicators, and the Consumer Price Index (CPI), which measures the average change in the prices of goods and services purchased by consumers and serves as a standard measure of inflation. The GSCPI data were obtained from the Federal Reserve Bank of New York~\cite{nyfed_gscpi}, and the CPI data from the Federal Reserve Economic Data (FRED) database~\cite{fred_cpiaucsl}. For regression analysis, we use the maximum value of GSCPI and the standard deviation of CPI. 

\subsection{Return cross-correlation matrix: construction and spectral decomposition}
\label{subsec:construction_cross_corr_eigendecomposition}
To construct the cross-correlation matrix that quantifies the linear correlations between the price movements of different stocks, we first convert the price series into logarithmic returns. This ensures that the results are not affected by the scale of measurement. The logarithmic return $R_i(t)$ for $i$-th stock at time $t$ is defined over a time-interval $\Delta t$ as

\begin{equation}
\label{eq:log_ret}
R_i(t) = \log \left(\frac{p_i(t+\Delta t)}{p_i(t)}\right)
\end{equation} where $p_{i}(t)$ denotes the adjusted closing price of the stock $i$ at time $t$. Since we use daily adjusted closing stock price data, we set $ \Delta t= 1$ day. 

Further, log returns are normalized, $r_i(t)$, to take into account the varying levels of volatility of different stocks,
\begin{equation}
\label{eq:norm_logret}
r_i(t) = \frac{R_i(t) - \langle R_i \rangle}{\sigma_i}
\end{equation} where $\sigma_i = \sqrt{\langle R_i^2 \rangle - \langle R_i \rangle^2}$ is the standard deviation of $R_{i}$, $\langle \cdot \rangle$ denotes the time average evaluated over the corresponding analysis window.

The interaction network is then constructed for each window by computing an equal-time cross-correlation matrix $C$, whose elements are defined as~\cite{plerou2002random}
\begin{equation}
C_{ij} = \left\langle r_i(t)\, r_j(t) \right\rangle
\label{eq:cross_corr}
\end{equation}

To filter out the non-random properties present in empirical correlation matrices, we compare the eigenvalue spectrum of $C$ with those of a random cross-correlation matrix~\cite{laloux1999noise}. The majority of eigenvalues of an empirical correlation matrix are consistent with those expected from a random correlation matrix, i.e., the Wishart matrix of size $N\times N$ constructed from $N$ mutually uncorrelated return time series, each of length $T$. The Marchenko--Pastur theorem describes the theoretical eigenvalue distribution for a random correlation matrix~\cite{marvcenko1967distribution}. In the limit $N \to \infty$, $T \to \infty$, such that $\frac{T}{N} >1$, the probability density function $P(\lambda)$ of eigenvalues $\lambda$ is given by

\begin{equation}
P(\lambda) =
\frac{T/N}{2\pi}
\frac{\sqrt{(\lambda_{+}-\lambda)(\lambda-\lambda_{-})}}{\lambda}
\end{equation}

for eigenvalues within the bounds $\lambda_{-} \leq \lambda \leq \lambda_{+}$, where 
\begin{equation}
\lambda_{\pm} = \left(1 \pm \sqrt{\frac{1}{T/N}}\right)^2
\end{equation}

Although the classical Marchenko--Pastur (MP) theorem assumes $T/N>1$, our empirical analysis is performed in the regime $T<N$. In this regime, the empirical correlation matrix is singular and belongs to the anti-Wishart ensemble, possessing exactly $N-T$ zero eigenvalues, which raises the question of whether the extraction of the global and group modes is obscured. However, RMT dictates that the MP distribution in this regime simply acquires a Dirac-delta function at zero to account for the singular eigenvalues, while the theoretical upper bound of the bulk spectrum ($\lambda_+$) remains analytically well-defined~\cite{pharasi2019complex}. Furthermore, it has been shown analytically that the joint probability distribution of the non-zero eigenvalues in the anti-Wishart case is exactly identical to that of a Wishart matrix, with the parameters $N$ and $T$ interchanged~\cite{vivo2007large}. This suggests the global and group modes that exceed the MP upper bound are separated from the degeneracy of zero eigenvalues at the bottom of the spectrum. Additionally, the probability of small fluctuations of the largest eigenvalue is governed by the Tracy-Widom distribution over a narrow width of $O(N^{1/3})$~\cite{vivo2007large}. Consequently, the MP bound remains applicable for identifying the informative eigenmodes.

The eigenvalues that lie within the bounds $\lambda_{\pm}$ constitute the bulk spectrum and are typically attributed to random noise. Eigenvalues exceeding $\lambda_{+}$ indicate genuine market structure and reflects interaction structure of group of stocks correlated in a similar fashion. Consequently, the empirical correlation matrix $C$ can be decomposed into three components ~\cite{kim2005systematic}:

\begin{equation}
C = C^{\text{global}} + C^{\text{group}} + C^{\text{random}}
\end{equation}

The global component $C^{\text{global}}$ constructed from the largest eigenvalue represents signals common to the entire market, where all stocks move collectively, 

\begin{equation}
C^{\text{global}} = \lambda_1\, v_1 v_1^{T}
\end{equation}

The random component $C^{\text{random}}$ is composed of the eigenvalues in the bulk of the spectrum.

\begin{equation}
C^{\text{random}} =
\sum_{\lambda_- \le \lambda_i \le \lambda_+}
\lambda_i\, v_i v_i^{T}
\end{equation}

Finally, the deviating eigenvalues (excluding the global mode), $\lambda_+ < \lambda_i < \lambda_1$, capture the mesoscopic structure of the market, by filtering out the market-wide movement (global mode) and the background noisy correlations (random mode), thereby revealing the group-specific correlations that constitute the primary focus of our work. Thus, the group mode can be reconstructed as,

\begin{equation}
C^{\text{group}} = \sum_{\lambda_+ < \lambda_i < \lambda_1} \lambda_i\, v_i v_i^{T}
\end{equation}

\subsection{Quantification of structural balance.}
\label{subsec:sbt_quantified} 
Following the RMT filtering, we reconstruct undirected signed networks from the group correlation matrices for each of the 56 periods. We apply a thresholding procedure to retain only the statistically significant connections~\cite{kuyyamudi2019emergence}. The adjacency matrix of the signed network is then defined as
\begin{equation}
A_{ij} =
\begin{cases}
-1, & \text{if } C_{ij}^{\text{group}} < \mu_{\text{random}} - 3\sigma_{\text{random}}, \\[6pt]
0,  & \text{otherwise}, \\[6pt]
1,  & \text{if } C_{ij}^{\text{group}} > \mu_{\text{random}} + 3\sigma_{\text{random}} .
\end{cases}
\label{eq:thresholding}
\end{equation}

where $\mu_{\text{random}}$ and $\sigma_{\text{random}}$ denote the mean and standard deviation of the off-diagonal entries of the random correlation matrix $C^{\text{random}}$ computed independently for each temporal window. The resulting thresholded signed networks provide the basis for quantifying structural balance over time.

Several quantitative measures have been developed to measure the degree of structural balance in signed networks~\cite{diaz2025signed}, including measures defined at multiple scales and those that account for the directed nature of interactions~\cite{aref2020multilevel}. Motif-based counting is one of the simplest approaches and can serve as a proxy to determine the balanced state of the overall network structure. In particular, the number of frustrated triads quantifies the degree of structural imbalance in the network and has been associated with the periods of systemic risk~\cite{kuyyamudi2019emergence}. However, merely counting frustrated triads is vulnerable to changes in network density across temporal windows. To overcome this limitation, we adopt the measure of polarization proposed by Pham et al.,~\cite{minh2020effect}. Derived from statistical mechanics, it serves as a normalized order parameter to characterize phase transitions in complex systems. It compares the fractions of balanced and unbalanced triads under the assumption that the aggregation of triad-level balance provides a meaningful representation of network-level balance~\cite{aref2020multilevel}. The formulation perfectly captures the theme of balance theory, whereby entities within a cohesive cluster form positive connections while connections between distinct groups tend to be negative. Importantly, the measure enables us to decompose the polarization at the global level into components based on the sectoral composition of triads. Accordingly, Polarization ($P$) is defined as the difference of fractions of balanced and unbalanced triads in the network.

\begin{equation} 
P = \frac{N_+ - N_-}{N_+ + N_-} 
\end{equation} 

where $N_+$ and $N_-$ represent the total number of balanced and unbalanced triads in the network, respectively. The metric is bounded within the interval $[-1, 1]$. All triads are balanced when $P=1$,  whereas $P<1$ indicates the presence of unbalanced triads, a state of structural imbalance.  A value of $P=0$ denotes an exact equilibrium between balanced and unbalanced triads, and $P<0$ indicates dominance of unbalanced triads. We evaluated the polarization on the adjacency matrices of the signed network at three distinct scales: globally across all the triads in the complete network $(P_{G})$, internally within individual sectors $(P^{intra})$, and across cross-sector triadic formations $(P^{inter})$. Throughout the study, the subscript $G$ indicates quantities computed over the entire network, whereas the superscripts $intra$ and $inter$ indicate quantities computed from within-sector and between-sector triads, respectively.

\begin{align} 
\text{Global polarization:} \qquad 
& P_G = \frac{N_+ - N_-}{N_+ + N_-} 
\label{eq:pg} 
\end{align} 

\begin{align} 
\text{Intra-sector polarization:} \qquad 
& P^{intra} =\frac{(N_+ - N_-)^{intra}}{(N_+ + N_-)^{intra}} 
\label{eq:p_intra} 
\end{align} 

\begin{align} 
\text{Inter-sector polarization:} \qquad 
& P^{inter} =\frac{(N_+ - N_-)^{inter}}{(N_+ + N_-)^{inter}} 
\label{eq:p_inter}
\end{align} 

The classification of global polarization into intra-sector and inter-sector categories is based on the sector labels of the constituent stocks forming the closed triads. Accordingly, the total number of balanced and frustrated triads can be decomposed as $N_{+}=N_{+}^{{intra}}+N_{+}^{{inter}}$ and $N_{-}=N_{-}^{{intra}}+N_{-}^{{inter}}$, respectively. Intra-sector triads consist exclusively of stocks belonging to the same sector, whereas an inter-sector triad contains stocks from two or more sectors.

\section{Results}

\subsection{Spectral analysis and structure of dominant eigenvectors}

\begin{figure}[!htbp]
    \centering
    \includegraphics[width=0.7\textwidth]{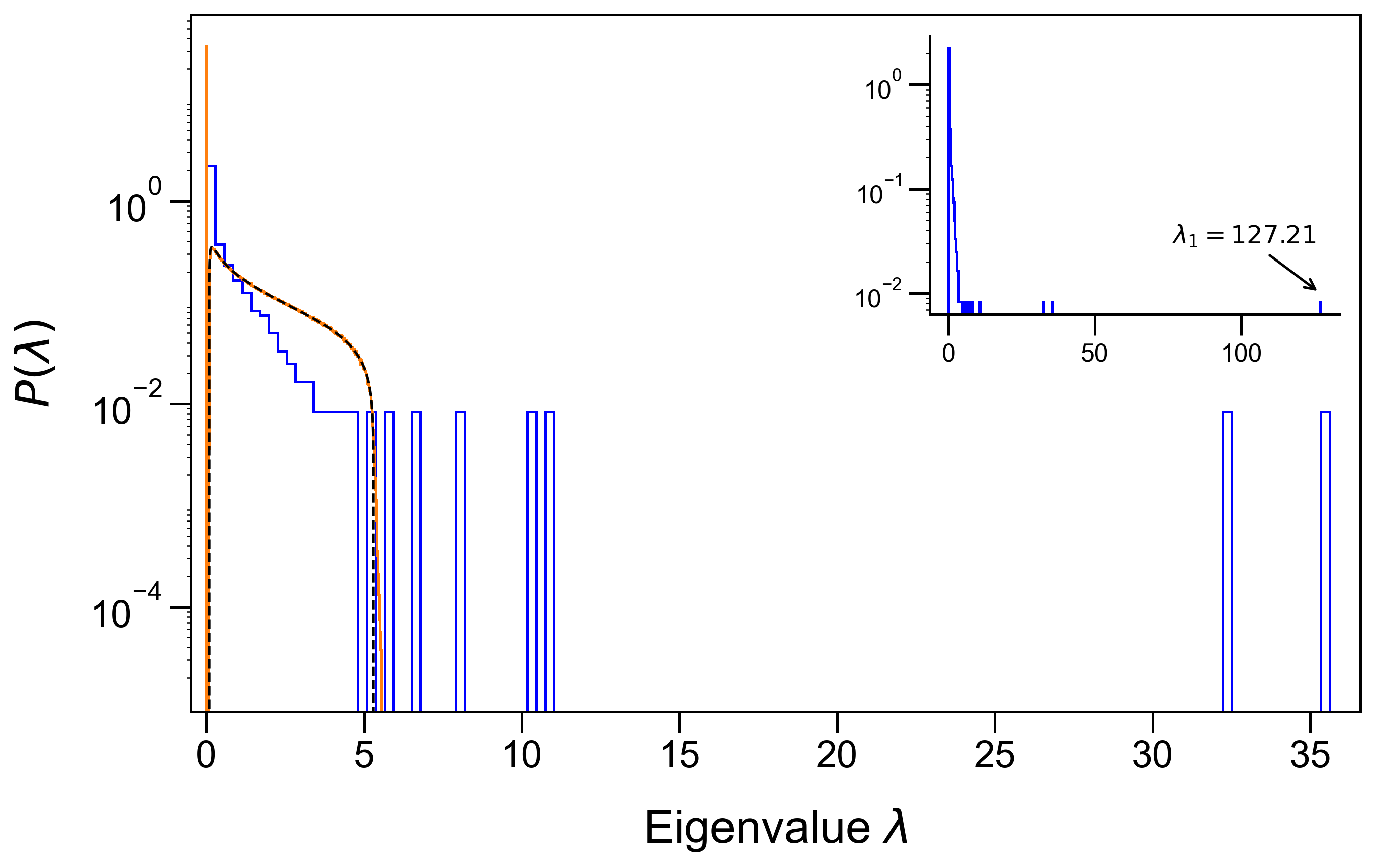}
    \caption{\textbf{Comparison of the eigenvalue spectra of the empirical correlation matrix (blue histogram), the Marchenko--Pastur (MP) distribution (dashed black curve), and the randomized (shuffled) correlation matrix (orange histogram) for a window during the onset of the COVID-19 crisis (dated April 09, 2021--April 06, 2022).} The insets display the complete eigenvalue spectrum, highlighting the 8 deviating eigenmodes, including the dominant market mode. The largest eigenvalue $\lambda_1$, corresponding to the market mode, is located far from the bulk of the eigenspectra, approximately 24 times larger than the MP upper bound $(\lambda_+ = 5.302)$.}
    \label{fig:spectral_density_period040}
\end{figure}

We begin by assessing whether the empirical correlation matrix contains a meaningful signal. Our analysis is based on shorter epochs of $T = 252$ days and $N = 426$ stocks, which remain consistent across all 56 windows; as a result, the aspect ratio $T/N = 0.59$. Under these conditions, we expect a broadening of the MP bulk, making it more difficult for empirical eigenvalues to stand out from the noise. To examine this, we compare the spectral density of the empirical correlation matrix with the corresponding MP distribution and the eigenvalue spectrum obtained from randomized (shuffled) correlation matrices. Among the 56 empirical correlation matrices, we present a representative example corresponding to a period during the COVID-19 crisis (April 09, 2021--April 06, 2022) in \autoref{fig:spectral_density_period040}. As discussed in \ref{subsec:construction_cross_corr_eigendecomposition}, the empirical correlation matrix falls under the anti-Wishart condition, resulting in a singular matrix. Therefore, its eigenvalue distribution contains $N-T$ zero eigenvalues. Despite this singularity, the majority of empirical eigenvalues lie within the MP bulk, while a few of the largest eigenvalues deviate significantly from the bulk. To ensure that these outliers are not a consequence of the finite length of the time series, we construct surrogate correlation matrices by independently randomizing the empirical return time series for each stock, producing a null model that destroys cross-correlation between assets. The corresponding eigenvalue distribution is estimated from 10,000 independent randomization trials. We observe a close correspondence between the eigenvalue distribution of these surrogate matrices and the MP prediction. Because the bulk of our empirical eigenvalue spectrum closely aligns with the eigenspectra obtained from surrogate matrices, we can conclude that the empirical bulk is consistent with the properties of a random correlation matrix spectrum. Although the surrogate eigenspectra exhibit a small number of eigenvalues that extend slightly beyond the MP upper bound because of finite-sample fluctuations, no pronounced outliers comparable to those in the empirical spectrum are observed. The surrogate matrices also exhibit the same zero degeneracy, as they preserve the dimensionality of the empirical correlation matrix. Overall, the close agreement between the empirical bulk, surrogate eigenspectra, and $P(\lambda)$ indicates that most of the empirical eigenvalue spectrum can be attributed to measurement noise, whereas the deviating eigenvalues represent genuine market correlations. This validates that the global and group modes extracted from our empirical matrices are not obscured by the anti-Wishart condition and genuinely reflect the market’s underlying structure. 

\begin{figure}[!htbp]
    \centering
    \includegraphics[width=\textwidth]{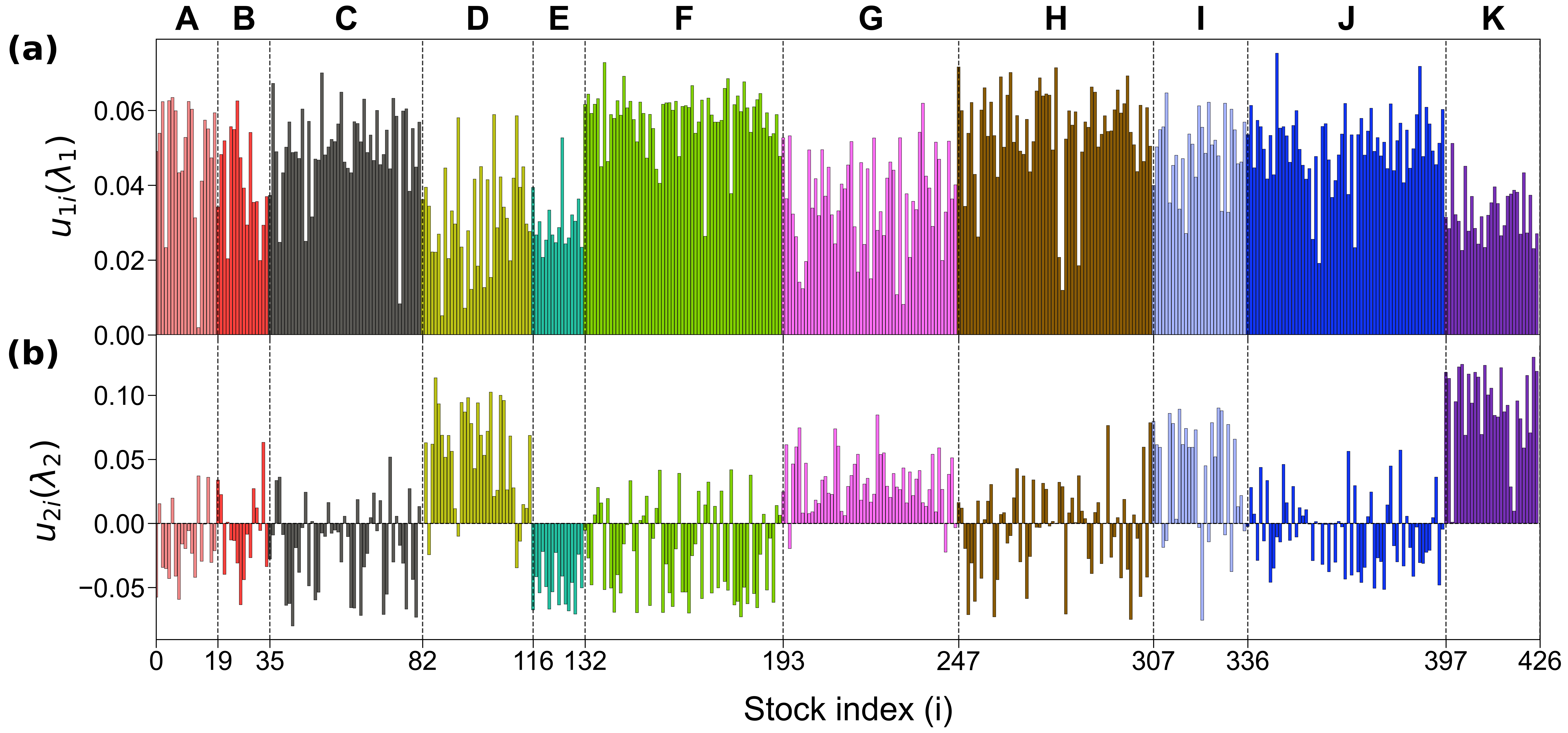}
    \caption{\textbf{Structure of the largest $(u_{1i})$ and second-largest eigenvector components $(u_{2i})$  of the S\&P 500 cross-correlation matrix $C$} during the period April 09, 2021--April 06, 2022. Panel (a) displays the eigenvector associated with the largest eigenvalue, and panel (b) displays that associated with the second-largest eigenvalue. Colors denote the sector affiliation of the constituent stocks. A: Basic Materials. B: Communication Services. C: Consumer Cyclical. D: Consumer Defensive. E: Energy. F: Financial Services. G: Healthcare. H: Industrials. I: Real Estate. J: Technology. K: Utilities.}
    \label{fig:structure_evect_compt}
\end{figure}

Having confirmed that these deviating eigenvalues represent genuine market signals, we now examine the structure of their associated eigenvectors to understand the contribution of individual stocks to the market structure and the sectoral interactions underlying the observed deviating eigenmodes. As a representative example among the 56 rolling windows, we present the eigenvector components corresponding to the largest and second-largest eigenvalues of the S\&P 500 cross-correlation matrix during a period of market instability (April 09, 2021--April 06, 2022). To examine the contribution of specific sectors to these dominant modes, we organize the stocks according to their industrial classification~\cite{kim2005systematic}. The eigenvector corresponding to the largest eigenvalue is shown in \autoref{fig:structure_evect_compt}(a). Its components display a relatively uniform composition with all elements exhibiting the same sign, reflecting the collective response of the entire market to external information. Hence, the largest eigenvalue corresponds to the market mode, indicating a mechanism that exerts a similar influence on all stocks. The representative example shown here qualitatively suggests that the market mode is delocalized. To quantify the degree of localization over all study periods we compute the inverse participation ratio (IPR)~\cite{plerou1999universal}, defined for the $k^{th}$ eigenvector as $I_{k} = \sum_{i=1}^{N} [u_{ki}]^4$, where $u_{ki}$ denotes the $i^{th}$ component of eigenvector $k$ and $k = 1$ corresponds to the market mode. In a completely delocalized eigenmode, where all components have identical contribution $(u_{ki} = 1/\sqrt{N})$, the IPR attains a minimum value $I_{1} = 1/N$. The average IPR of the market mode over the 56 rolling windows is $\langle I_{1} \rangle = 0.00280 \pm 0.00029$, which is close to the completely delocalized mode $1/N = 0.00235$ for $N=426$. The close proximity of these values indicates that the market mode remains delocalized throughout the study periods and involves the collective participation of nearly all stocks. We next examine the signed components of the second largest eigenvector~\cite{sinha2013uncovering}, shown in \autoref{fig:structure_evect_compt}(b), to reveal the underlying sectoral dynamics. In contrast to the uniform composition observed in market mode, the group mode exhibits significant variations in the signs of its components across sectors. For instance, sectors such as Energy and Technology display signs opposite to those of Healthcare and Utilities. The opposite signs in the components of an eigenvector correspond to anti-correlated movement between the corresponding sectors, indicating that some sectors generally move in opposite directions. Therefore, this establishes the genesis of the negative edges observed in the group structure extracted from the S\&P 500 cross-correlation matrix. 


\subsection{Sectoral origin of structural imbalance during systemic risk}
\label{sec:all_polarization_findings}
\begin{figure}[!htbp]
    \centering
    \includegraphics[width=\textwidth]{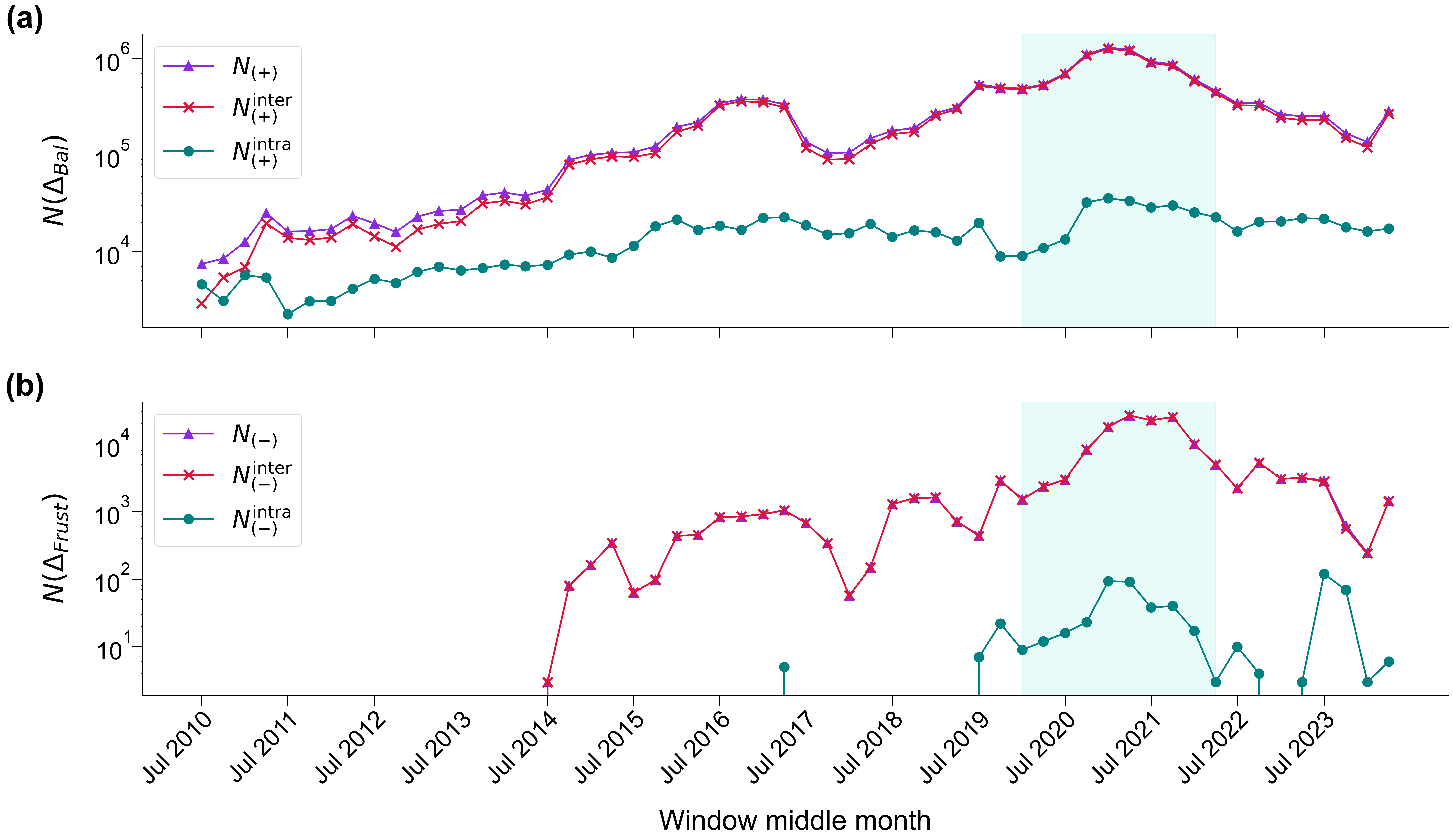}
    \caption{\textbf{Temporal evolution of balanced $N(\Delta_{Bal})$ and frustrated $N(\Delta_{Frust})$ triads, resolved into intra-sector and inter-sector configurations.}  (a) shows the total number of balanced triads, $N_{+}$, along with the numbers of inter-sector ($N_{+}^{{inter}}$) and intra-sector ($N_{+}^{{intra}}$) balanced triads. (b) shows the corresponding quantities for frustrated triads, namely the total number $N_{-}$ and its inter-sector ($N_{-}^{{inter}}$) and intra-sector ($N_{-}^{{intra}}$) components. The shaded region denotes the period associated with the COVID-19 pandemic, spanning from the window middle date, Jan 08, 2020, to April 04, 2022.}
    \label{fig:bal_frust_sector_triads_evol}
\end{figure}

Given that these anti-correlated movements generate negative edges within the empirical group structure, we now explore the competing interactions these edges introduce into the network, which are especially heightened during periods of systemic risk. To investigate how this conflicting configuration alters the network's structural organization, we analyze the adjacency matrices of the signed network by employing the polarization metrics proposed in \autoref{subsec:sbt_quantified}. Because our aim is to quantify how the structural imbalance is distributed within and across sectors, we track the temporal evolution of the inter- and intra-sector balanced and frustrated triads over the study period. The temporal evolution of the number of balanced triads in the signed network of S\&P 500, shown in \autoref{fig:bal_frust_sector_triads_evol}(a), decomposed according to their sectoral composition, reveals that balanced triads of all categories are persistently present and substantially more numerous than frustrated triads throughout the observation period. Moreover, the number of inter-sector balanced triads ($N_{+}^{{inter}}$) generally exceeds the number of intra-sector balanced triads ($N_{+}^{{intra}}$), with the exception of the initial stage of the observation period, where an opposite trend is observed. In contrast, the temporal evolution of frustrated triads shown in \autoref{fig:bal_frust_sector_triads_evol}(b) does not show the presence of frustrated triads until the beginning of the third quarter of 2014, which is then followed by a gradual increase in the number of inter-sector frustrated triads ($N_{-}^{{intra}}$), and remains elevated during the COVID-19 pandemic. A particularly noteworthy feature is the emergence of intra-sector frustrated triads ($N_{-}^{{intra}}$), which become especially pronounced during the COVID-19 crisis event and persist into the post-crisis recovery phase. This behavior reveals structural information that is not readily apparent from the aggregate evolution of frustrated triads alone. For comparison, we examine the temporal evolution of inter-and intra-sector balanced and frustrated triads during the 2008 GFC in SI Text S1.2. Contrary to the finding reported for the COVID-19 period, we find that the 2008 GFC exhibits a complete absence of the intra-sector frustrated triads. Taken together, these observations suggest that intra-sector frustrated triads constitute a more distinctive signature of the market instability induced by the COVID-19 crisis than of the 2008 GFC, and may encode information about the nature of the underlying source of systemic disruptions. 

\begin{figure}[!t]
    \centering
    \includegraphics[width=\textwidth]{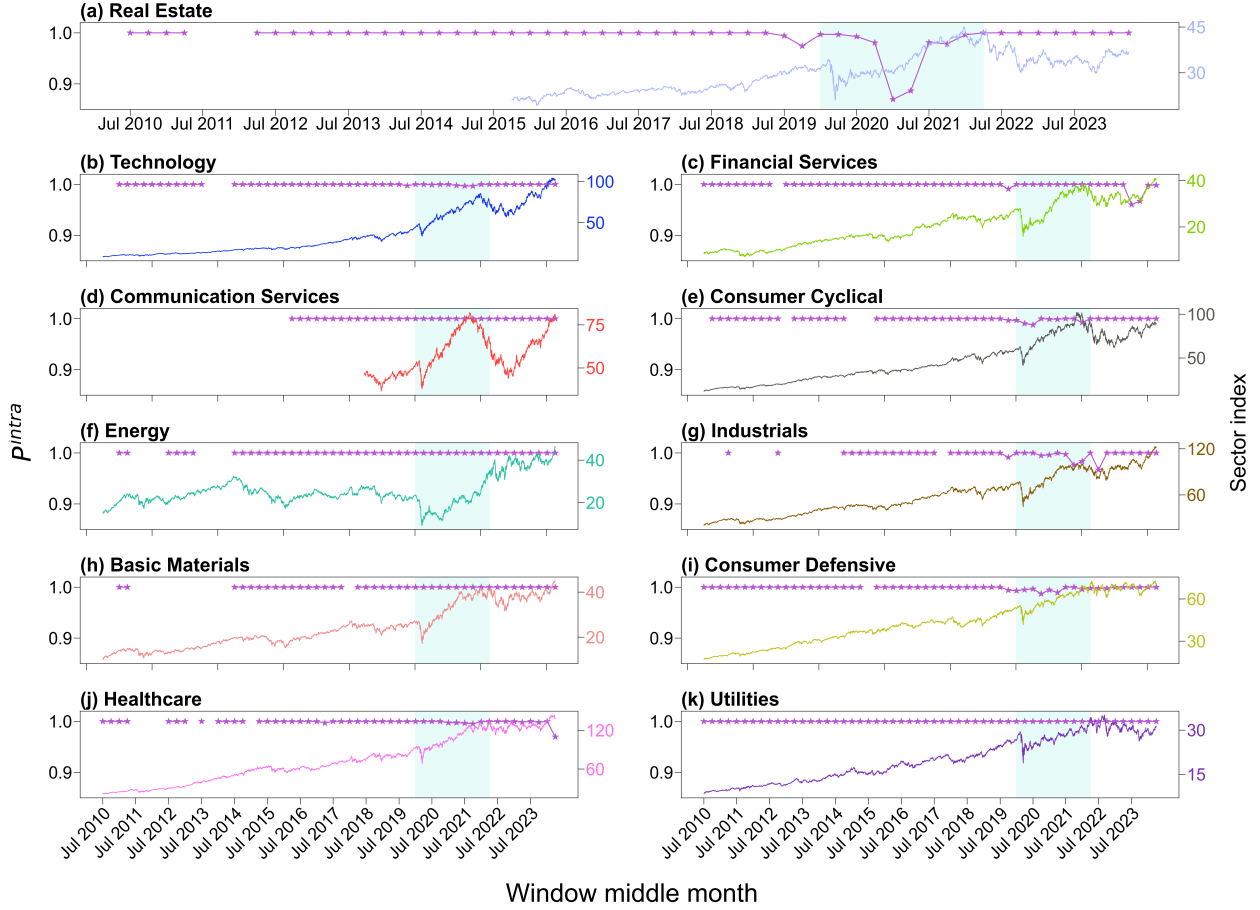}
    \caption{\textbf{Temporal evolution of intra-sector polarization across the 11 GICS sectors.} (a)–(k) present $P^{intra}$, for each of the 11 GICS sectors in the S\&P 500 over the study period. The corresponding sector index is shown on the secondary axis in each subplot. The shaded area indicates the COVID-19 region. Missing values of $P^{intra}$ correspond to periods in which no triads were formed within the sector, rendering the polarization measure undefined. The Real Estate and Communication Services sector indices appear only from their respective inception dates, following their establishment as standalone GICS sectors in 2016 and 2018, respectively.}
    \label{fig:p_intra}
\end{figure}

We now shift our focus from triad-level analysis to the mesoscale to investigate the distribution of frustration within and across sectors. By examining the network's internal sectoral organization, we track the intra-sector polarization $(P^{intra})$ across distinct market domains. For each temporal network, we extract the sector-specific subgraphs from the signed adjacency matrix, and the number of balanced and unbalanced triads is computed to evaluate $P^{intra}$. The temporal evolution of $P^{intra}$ for all sectors is shown in \autoref{fig:p_intra}. A notable feature of the results is the persistence of high intra-sector polarization across the study periods. During the periods of market stability, many sectors exhibit $P^{intra}=1$, indicating the complete absence of frustrated triads within sector boundaries. Temporary reductions in $P^{intra}$ are observed during the periods of market instability in sectors such as Real Estate, Industrials, Consumer Cyclical, and Consumer Defensive. Among these, Real Estate exhibits a prominent decline between late 2020 and late 2021 due to the opposing movement of stocks within the sector. This behavior is likely associated with the heterogeneous impact of the pandemic across different industries within the real estate sector. Commercial real estate sector, hotel and retail properties suffered a severe decline due to lockdown, whereas businesses operating at industrial production sites were less affected~\cite{balemi2021covid}. Despite these temporary reductions, crisis periods continue to exhibit strongly balanced intra-sector structures even when their sector indices show substantial declines. This event can largely be attributed to a synchronized market collapse. When stocks within a sector plummet simultaneously, their movements become highly correlated, resulting in predominantly positive relationships that give rise to balanced triadic structures despite overall negative performance. Overall, the mesoscopic analysis reveals that, regardless of the market's economic situation, frustration remains weak within sector boundaries. The corresponding analysis for the 2008 GFC presented in SI Figure S2 exhibits qualitatively similar behavior.

\begin{figure}[!htbp]
    \centering
    \includegraphics[width=\textwidth]{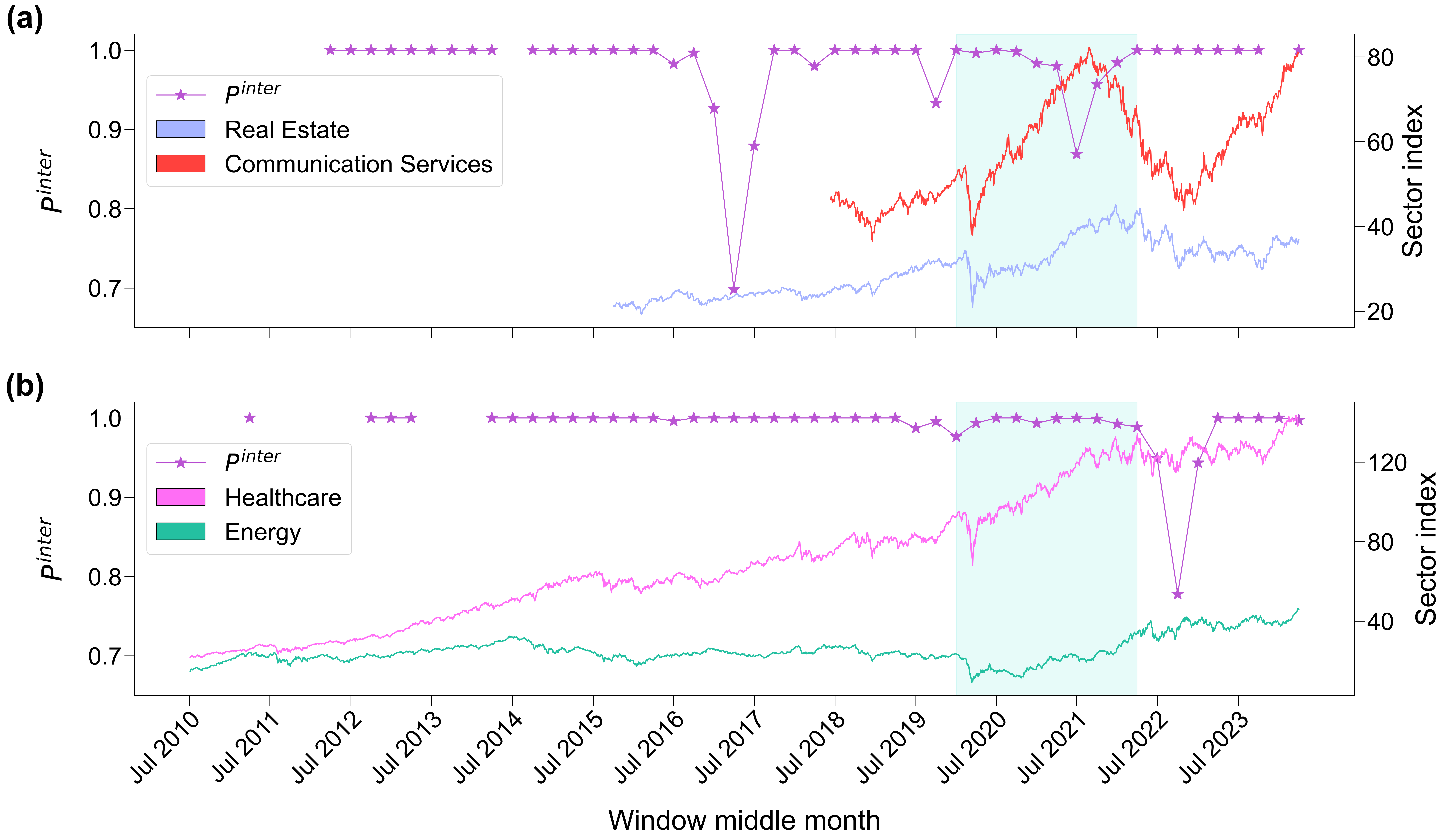}
    \caption{\textbf{Temporal evolution of inter-sector polarization $(P^{inter})$ for the two least-polarized sector pairs.} Sector indices are shown on the secondary axis, and the shaded area indicates the COVID-19 region. $P^{inter}$ is undefined during periods in which no inter-sector triads are formed, resulting in gaps in the curves. The Real Estate and Communication Services sector indices appear only from their respective inception dates, following their establishment as standalone GICS sectors in 2016 and 2018, respectively.}
    \label{fig:p_inter}
\end{figure}

We further extend the mesoscopic analysis beyond individual sectors to examine the interactions between sectors. The temporal evolution of inter-sector polarization $(P^{inter})$ is presented in \autoref{fig:p_inter}. The polarization between sectors is computed on combined sector subgraphs by examining cross-sector triads that consist of two nodes from one sector and one node from the other. Because the S\&P 500 contains 11 GICS sectors, pairwise analysis yields 55 sector combinations. For clarity, we focus on the sector pairs exhibiting the lowest values of $P^{inter}$ throughout the observation period. In this case, the minimum inter-sector polarization is observed between the Real Estate and Communication Services sectors from late 2016 to late 2017, followed by reductions pre- and during the COVID-19 crisis, as shown in \autoref{fig:p_inter}(a). The Healthcare--Energy sector pair exhibits the second-lowest value of $P^{inter}$, particularly in the early 2022 to 2023 window, as shown in \autoref{fig:p_inter}(b). In contrast to the strong balance observed within sectors, a remarkable pattern emerges in which the sectors that acted as an indivisible block over the periods under study now show substantially lower polarization, due to the incompatibility of relationships across sectors. An analogous pattern is observed for the 2008 GFC dataset (see SI Figure S3).

To inspect the drivers behind the pronounced declines in $P^{inter}$ during the identified periods, we identify all the frustrated triads emerging within the corresponding window, then examine the frequency with which each pair of stocks forms the negative edge in these triads. To ensure these negative relationships are not caused by overall market movement, we compute the relative partial correlation $(\rho_{i,j:SP})$ for each pair while controlling for the S\&P 500 index.~\cite{baba2004partial, kenett2010dominating}. Here, we have discussed the cases that exhibit the highest frequency among the identified inter-sector negative edge pairs. A comprehensive list of the five highest-frequency inter-sector negative edges is provided in SI Text S2. For both representative sector pairs shown in \autoref{fig:p_inter}, the reduction in $P^{inter}$ is entirely associated with the predominance of $(++-)$ triads.  This imbalance arises when an inter-sectoral triad consists of an intra-sector edge that is positive, alongside one inter-sector positive edge and one negative edge, thereby creating a structural imbalance. For instance, consider the window spanning Oct 2016 to Oct 2017, the Real Estate--Communication Services sector pair attains its minimum value of $P^{inter}=0.69$. Among the $(++-)$ triads identified, the stock pairs (GOOGL, KIM) $(\rho_{\text{GOOGL,KIM:SP}}=-0.27)$, (GOOGL, SPG) $(\rho_{\text{GOOGL,KIM:SP}}=-0.21)$ appear to be the most frequent inter-sector negative edges, all of which exhibit negative partial correlation. A similar, although less pronounced, decline occurs during the COVID-19 pandemic (early 2021 to 2022), when $P^{inter}$ decreases to 0.86. The extraction of negative edge reveals a high frequency of pairs such as (EXR, LYV) $(\rho_{\text{EXR,LYV:SP}}=-0.11)$, (LYV, PSA) $(\rho_{\text{LYV,PSA:SP}}=-0.21)$, which majorly contributed to the frustration during this period. Similar analysis over the Healthcare--Energy sector pair over the window spanning early 2022 to 2023, during which $P^{inter}$ falls to 0.77, identifies stock pairs such as (COP, JNJ) $(\rho_{\text{COP,JNJ:SP}}=-0.16)$, (EOG, JNJ) $(\rho_{\text{EOG,JNJ:SP}}=-0.17)$ largely contributing to the high frequency of inter-sector negative edges. These pairs consistently exhibit negative partial correlations, highlighting their contribution to the observed frustration. Collectively, these findings demonstrate that structural imbalance in the market must reside at the boundaries between fundamentally opposed sectors.

\begin{figure}[!htbp]
    \centering
    \includegraphics[width=\textwidth]{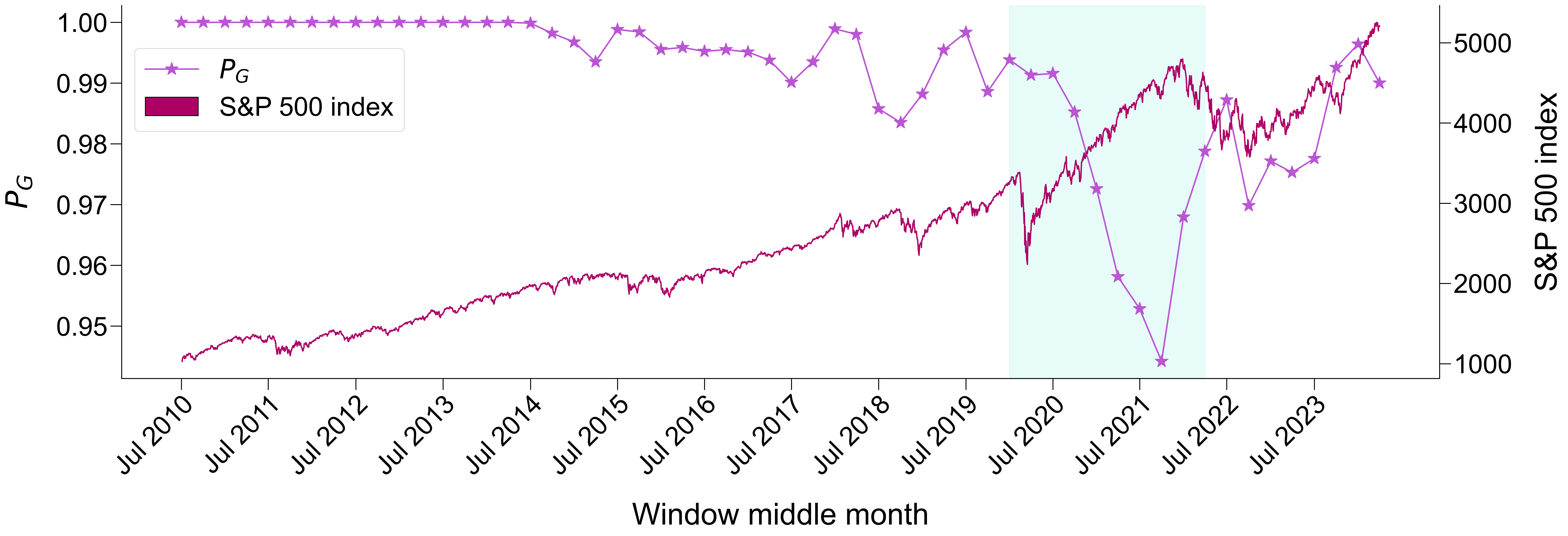}
    \caption{\textbf{Temporal evolution of global polarization.} The S\&P 500 index is shown on the secondary y-axis; the shaded area denotes the COVID-19 region.}
    \label{fig:pg}
\end{figure}
Having examined polarization both within and across sectors, we finally investigate the network at the macroscopic level through the global polarization $(P_G)$, whose temporal evolution is shown in \autoref{fig:pg}. From 2010 to 2015, the market exhibited a structurally balanced regime $(P_G = 1)$. This period also coincides with a sustained upward trajectory of the S\&P 500. Following 2015, small fluctuations begin to appear in $P_G$, indicating the emergence of localized structural inconsistencies within the network. These fluctuations coincide with several economic disruptions, including the 2015-2016 stock market selloff~\cite{kang2025indicator}, the 2018 Volmageddon, and the Repo Crisis 2019~\cite{hammond2026geometric}. The most pronounced departure from structural balance occurs during the COVID-19 crisis, highlighted by the shaded region in \autoref{fig:pg}. The systemic risk period during the COVID-19 pandemic caused severe, historic volatility for the S\&P 500. During this period, $P_G$ experienced a sharp decline, reflecting a substantial increase in the number of unbalanced triads relative to balanced ones. Since intra-sector structures remain largely balanced throughout the study period, the reduction in $P_G$ must predominantly arise from the accumulation of frustrated configurations across sector boundaries. The corresponding temporal evolution of $P_G$ for the 2008 GFC dataset is provided in SI Figure S4.

\subsection{Decomposing global polarization into intra- and inter-sector contributions}

The preceding analysis of cross-sector polarization leads to the conclusion that the loss of global structural balance must predominantly emerge from interactions between different sectors rather than within them. This motivates us to decompose the global polarization metric to quantify the specific contributions of intra-sector and inter-sector components to the overall structural balance across the mesoscale structure of the S\&P 500 network.  

As discussed in section~\ref{subsec:sbt_quantified}, the total number of balanced $N_+$ and frustrated $N_-$ triads in the network is the sum of their respective intra-sector and inter-sector components. The net structural balance of the network can therefore be expressed as the sum of these differences. Normalizing by the total number of triads  ($N=N_+ + N_-$) in the network yields:  

\begin{equation}\frac{N_+ - N_-}{N} = \frac{(N_+ - N_-)_{G}^{intra}}{N} + \frac{(N_+ - N_-)_{G}^{inter}}{N}{N} 
\label{eq:norm_expansion} 
\end{equation} 

We now define the global intra-sector and inter-sector polarization metrics ($P_{G}^{intra}$ and $P_{G}^{inter}$), which are normalized by the total number of intra-sector ($(N_+ + N_-)_{G}^{intra}$) and inter-sector triads ($(N_+ + N_-)_{G}^{inter}$) in the complete network, respectively. Further, substituting these subset polarization metrics into ~\eqref{eq:norm_expansion} decomposes $P_G$. 

\begin{equation} 
P_{G} = P_{G}^{intra}W_{G}^{intra} + P_{G}^{inter}W_{G}^{inter} 
\label{eq:pg_decomposed} 
\end{equation} 

where $W_{G}^{intra/inter}$ denotes the fraction of the network's triads that belong to each category, i.e. 

\begin{equation} 
W_{G}^{intra/inter} = \frac{(N_+ + N_-)_{G}^{intra/inter}}{N} 
\end{equation} 

\begin{figure}[!htbp]
    \centering
    \includegraphics[width=\textwidth]{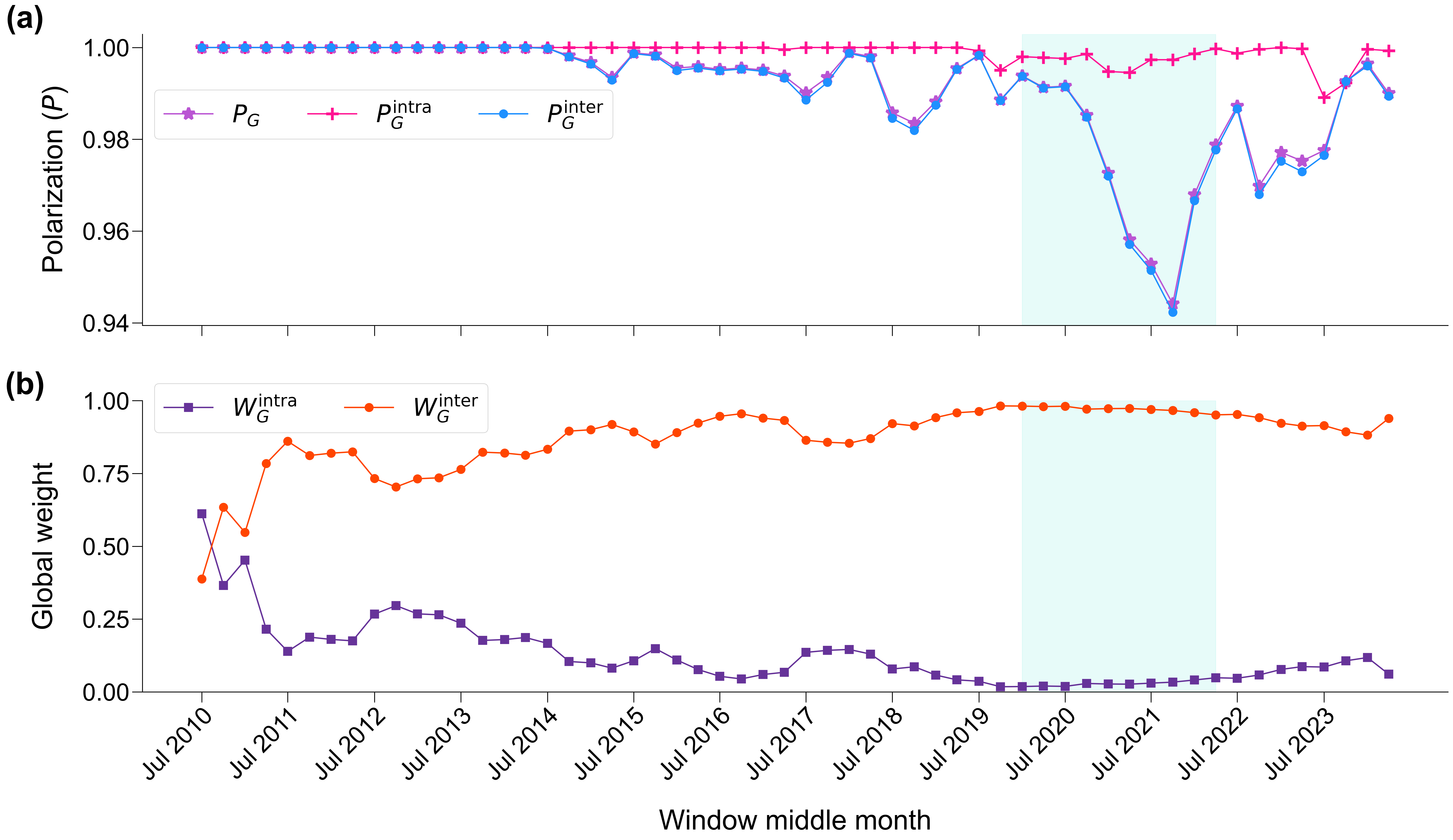}
    \caption{\textbf{Temporal evolution of global polarization components and their corresponding weights.} (a) shows the evolution of global polarization $(P_{G})$, global intra-sector polarization $(P_{G}^{intra})$, and global inter-sector polarization $(P_{G}^{inter})$. (b) shows the evolution of intra-sector weight $(W_{G}^{intra})$ and inter-sector weight $(W_{G}^{inter})$.}
    \label{fig:pg_decomposed_fig}
\end{figure}

Equation\eqref{eq:pg_decomposed} reveals that global polarization is fundamentally a weighted sum of the global intra-sector and inter-sector polarizations. This decomposition provides a natural framework for interpreting temporal variations in $P_G$ in terms of changes in the relative contributions of within-sector and between-sector interactions. Moreover, it enables us to view $P_G$ as a continuous measure that can be used to visualize the gradual decay or rise of structural imbalance in a complex system. \autoref{fig:pg_decomposed_fig} presents the temporal evolution of these contributing elements. The global inter-sector polarization closely mirrors the whole-network polarization, confirming that global structural imbalance is primarily driven by conflict between sectors rather than within sectors.  

Furthermore, we also examine the temporal evolution of the corresponding weights ($W_{G}^{intra}$ and $W_{G}^{inter}$) and observe that the intra-sector weight is strictly greater than the inter-sector weight during the first window, an effect of the large number of intra-sector balanced triads seen during the initial observation period (see \autoref{fig:bal_frust_sector_triads_evol}). However, this trend subsequently reverses, with inter-sector triads accounting for the larger fraction of the network throughout the remainder of the study period. Because these two weights represent fractions of the whole network, they inherently sum to 1, exhibiting perfectly symmetric behavior over time. 

Subsequently, a standard prerequisite for verifying structural balance is to benchmark the empirical metrics against a null model to ensure that the observed patterns do not arise purely by chance~\cite{kirkley2019balance}. Therefore, to assess whether the observed polarization patterns arise from the underlying structural constraints of the network rather than by chance, we first benchmark the empirical network against a signed degree-preserved randomization null model by utilizing a modified sign-constrained Maslov-Sneppen algorithm~\cite{maslov2002specificity}. The empirical graph is first partitioned into positive and negative subgraphs. A degree-preserved randomized version is generated for each subgraph by repeatedly selecting two edges at random and rewiring their endpoints. To preserve the simple signed topology of the financial market, the algorithm strictly rejects swaps that would introduce self-loops, multiple edges, or signed edge collisions (i.e., placing a positive edge where a negative edge already exists in the complementary subgraph). The randomized positive and negative subgraphs are subsequently recombined to produce the final null model. We quantify the statistical significance of each polarization metric using the z-score.
\begin{equation}
    z = \frac{P_{emp}-\langle P_{rand} \rangle}{\sigma(P_{rand})}
    \label{z_score_maslov_sneppen}
\end{equation}
where $P_{emp}$ is the empirical polarization metric, and $\langle P_{rand}\rangle$ and $\sigma(P_{rand})$ denote the mean and standard deviation of the corresponding metric computed over an ensemble of 1,000 independently generated signed degree-preserving null networks.

\begin{figure}[!htbp]
    \centering
    \includegraphics[width=\textwidth]{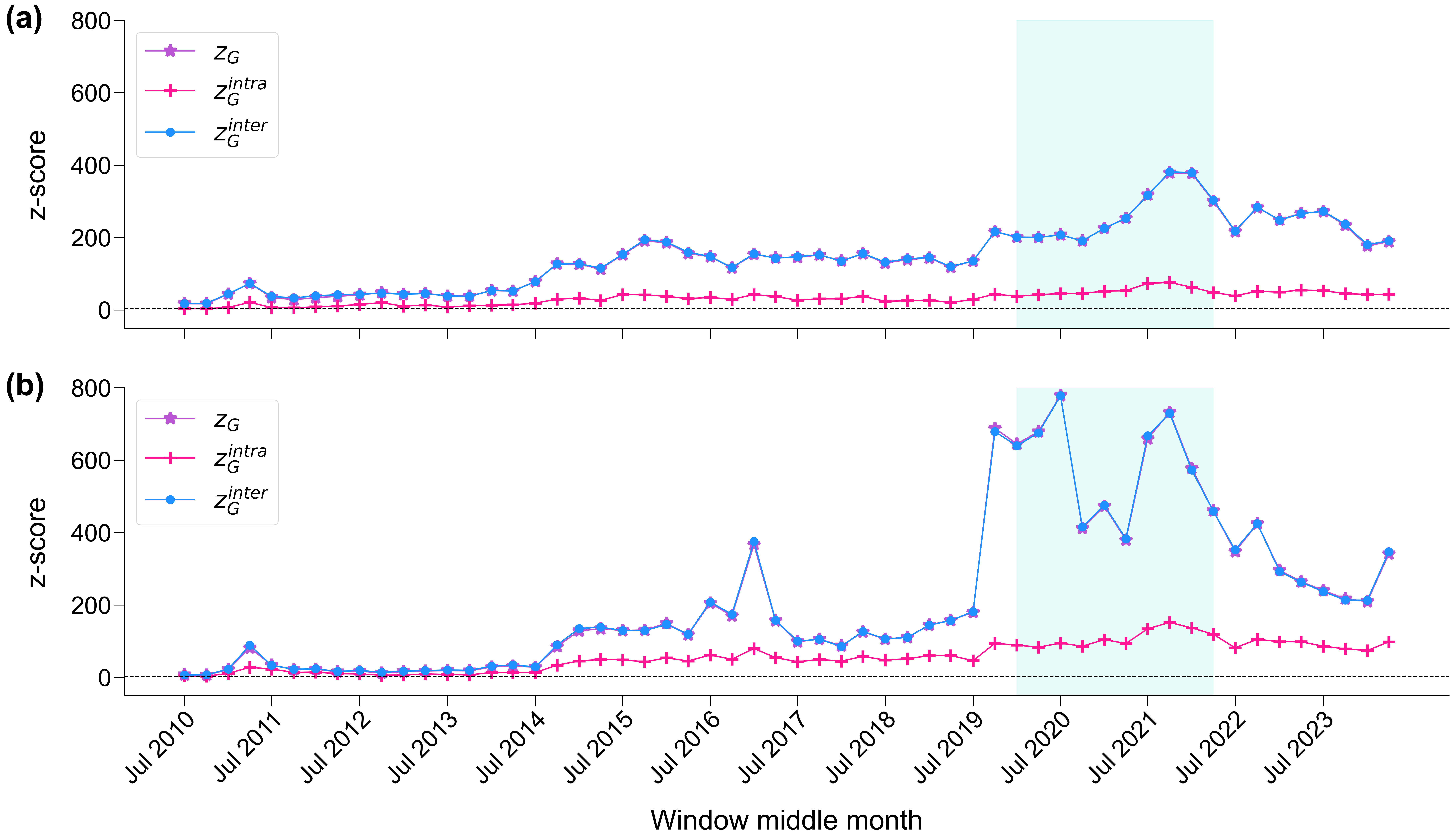}
    \caption{\textbf{Temporal evolution of the z-score for the global polarization $z_G$, global intra-sector polarization $z_{G}^{intra}$, and global inter-sector polarization $z_{G}^{inter}$} under (a) the sign-rewiring null model and (b) the signed topology-preserving (STP) maximum-entropy null model. The z-scores are computed independently for each polarization metric. The horizontal dotted line at $z=3$ marks the threshold for statistical significance, while the shaded region denotes the COVID-19 pandemic period.}
    \label{fig:z_score_nulls_covid}
\end{figure}

As illustrated in \autoref{fig:z_score_nulls_covid}(a), the z-scores for all polarization metrics ($P_G, P_G^{intra}, P_G^{inter}$) remain above the standard significance threshold ($z > 3$) across the entire 14-year timeline. This confirms that the empirical structural balance of the market deviates from random expectation. 

However, while benchmarking the empirical network against a conventional signed degree-preserved null model yields highly significant z-scores, the standard edge-rewiring protocol completely destroys the inherent topology in the empirical network. As recently demonstrated by Hao and Kov\'acs~\cite{hao2024proper}, traditional randomization methods can fail to detect balance even within perfectly balanced signed graphs due to the loss of local node heterogeneity and failing to preserve topology. To address this limitation and provide a rigorous test, we cross-validate the findings from the signed rewire null model using the signed topology-preserving (STP) null model based on the maximum entropy framework. We generated the STP null model using the original implementation provided by the authors~\cite{hao2024stpcode}. As shown in \autoref{fig:z_score_nulls_covid}(b), the STP null model produces qualitatively identical results. The z-scores for all constituents of global polarization remained definitively significant ($z > 3$) across the entire 14-year timeline. In SI Text S1.4, we perform a similar baseline comparison on the 2008 GFC dataset, which yields qualitatively consistent and statistically significant departures from the corresponding null models. The survival of our empirical findings under both the conventional rewiring protocol and the highly conservative STP baseline demonstrates that they are robust to the choice of null models, suggesting that these structural responses are consequences of economic stress rather than random network organization.

\FloatBarrier
\subsection{Regression analysis of global polarization}

To investigate the influence of macroeconomic indicators on global polarization $(P_G)$, we treat $P_G$ as the response variable and model its relationship with the two explanatory variables derived from the global supply chain pressure index (GSCPI) and the consumer price index (CPI). Both the macroeconomic indicators are available at a monthly resolution. To ensure temporal consistency with the 56 overlapping windows over which $P_G$ is computed, we aggregate both the measures by extracting the observations corresponding to the window start and end dates. Window-level summary statistics are then computed, with the maximum GSCPI $(\mathrm{GSCPI}_{\max})$ value capturing the peak supply chain disruption within each window and the standard deviation of CPI $(\mathrm{CPI}_{\mathrm{std}})$ quantifying inflation uncertainty. These two window-level summary statistics of the selected macroeconomic variables serve as the explanatory variables in the subsequent regression analysis.

\begin{figure}[!htbp]
    \centering
    \includegraphics[width=\textwidth]{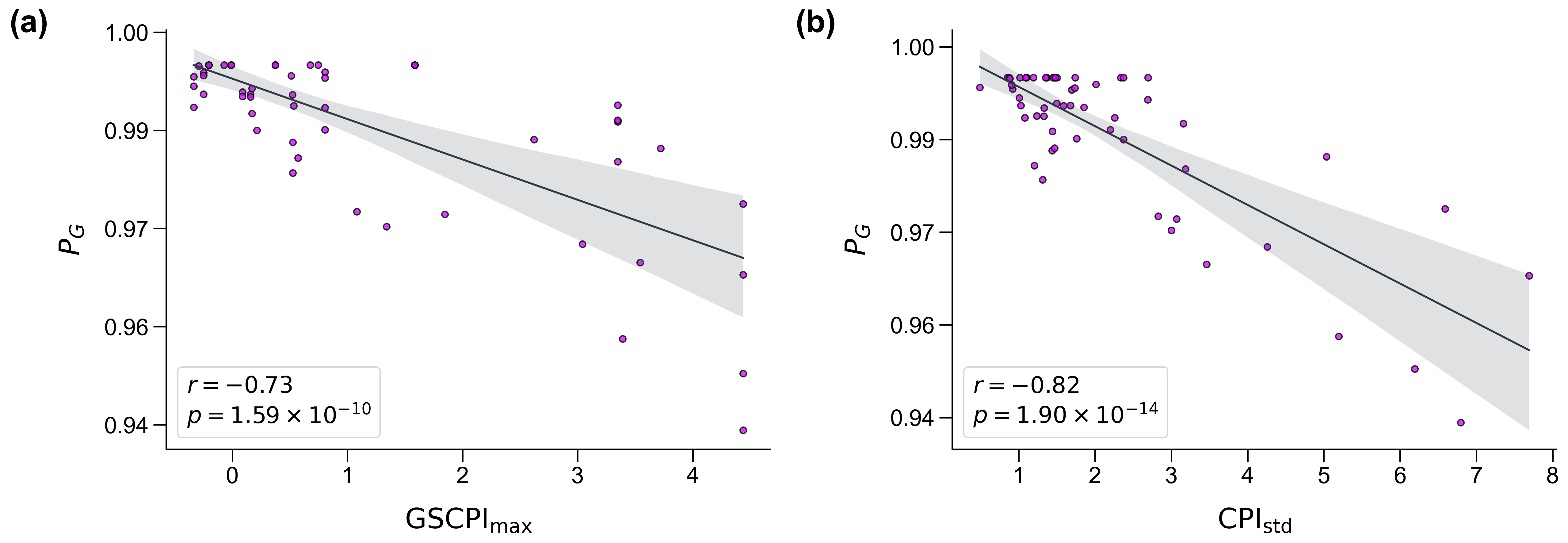}
    \caption{\textbf{Pairwise relationships between global polarization and macroeconomic variables.} (a)  maximum value of GSCPI $(\mathrm{GSCPI}_{\max})$ during the specific window and (b) the window-wise standard deviation of the CPI $(\mathrm{CPI}_{\mathrm{std}})$.}
    \label{fig:regplot_gscpi_cpi}
\end{figure}

\autoref{fig:regplot_gscpi_cpi} illustrates the bivariate relationship between $P_G$, $\mathrm{GSCPI}_{\max}$, and $\mathrm{CPI}_{\mathrm{std}}$, with a least-squares regression line. We observe that both variables exhibit a strong negative correlation with global polarization. The 95\% confidence interval widens slightly towards the extremes of the explanatory variables due to fewer observations. Both relationships exhibit approximately linear trends, providing further confirmation for the subsequent multiple linear regression.

To quantify the relationship between global polarization and the explanatory variables, we adopt a multiple linear regression model using ordinary least squares.  To ensure comparability of the estimated effects, both explanatory variables are z-score standardized (denoted by the tilde $\sim$). The resulting regression model is

\begin{equation}
\hat{P}_{G}
= 0.99071
- 0.00274\,\tilde{\operatorname{GSCPI}}_{\max}
- 0.00816\,\tilde{\operatorname{CPI}}_{\mathrm{std}}
\label{eq:regression}
\end{equation}

The regression statistics are reported in \autoref{tab:ols}. The estimated coefficients associated with $\mathrm{GSCPI}_{\max}$ and $\mathrm{CPI}_{\mathrm{std}}$ are negative, indicating that an increase in explanatory variables is associated with lower levels of global polarization. All regression coefficients are statistically significant at the 5\% significance level ($p < 0.05$). The regression model achieves a coefficient of determination of $R^2 = 0.682$ and an adjusted coefficient of determination of $\bar{R}^2 = 0.670$, indicating that approximately 68.2\% of the variation in global polarization is explained by the joint effects of the maximum of GSCPI and the standard deviation of CPI. The overall regression model was statistically significant $(F = 22.49 \text{ and } p = 8.49\times10^{-8})$. Collectively, these findings suggest that global polarization is associated with global supply chain disruption and inflationary uncertainty.

\vspace{10pt}
\begin{table}[!htbp]
\centering
\begin{threeparttable}
\normalsize
\renewcommand{\arraystretch}{1.25} 
\setlength{\tabcolsep}{10pt}        

\begin{tabular}{
    l
    S[table-format=1.6]
    S[table-format=1.6, group-digits=false]
    c
    S[table-format=1.6]
}
\toprule
{Feature} & {Coefficient} & {Std. error} & {$p$-value} & VIF \\
\midrule
$\mathrm{GSCPI}_{\max}$ & -0.00274 & 0.00096 & 0.00423 & 4.08516\\
$\mathrm{CPI}_{\mathrm{std}}$ & -0.00816 & 0.00201 & $5\times10^{-5}$ & 4.08516\\
\bottomrule
\end{tabular}
\caption{OLS regression coefficients for the standardized explanatory variables. Standard errors are Heteroskedasticity and Autocorrelation Consistent (HAC) to strictly account for the overlapping temporal windows. The last column reports the variance inflation factors (VIFs), computed prior to model estimation. Both explanatory variables exhibit VIF values of 4.09, below the commonly accepted threshold of 5, indicating no evidence of multicollinearity.}
\label{tab:ols}
\end{threeparttable}
\end{table}

\section{Discussion}

We have studied the interaction structure of the S\&P 500 through the lens of structural balance theory using a longitudinal dataset spanning 14 years from Jan 2010 to Dec 2024, covering a major economic crisis triggered by the COVID-19 outbreak. In this work, we have focused on the interactions among sectors to map the distribution of frustration in the complex financial system throughout periods of market stability and instability. Since the stocks of a sector of a financial market share common economic properties, it is interesting to observe how their interplay governs the structural balance throughout the periods under study. Because these underlying sectoral interactions are often obscured by global market trends, we utilized RMT filtering to isolate the meaningful group structure from the cross-correlation matrices.  Our initial analysis confirms that the emergence of negative edges in the group mode is driven by distinct anti-correlation between sectors. This behavior is clearly captured by the opposing signs in the second largest eigenvector of the cross-correlation matrix. Building upon the framework of ~\cite{kuyyamudi2019emergence}, we subsequently employ a triadic order parameter~\cite{minh2020effect, pham2021balance} to track the critical transition that the financial market undergoes.  

\begin{figure}[!htbp]
    \centering
    \includegraphics[width=\textwidth]{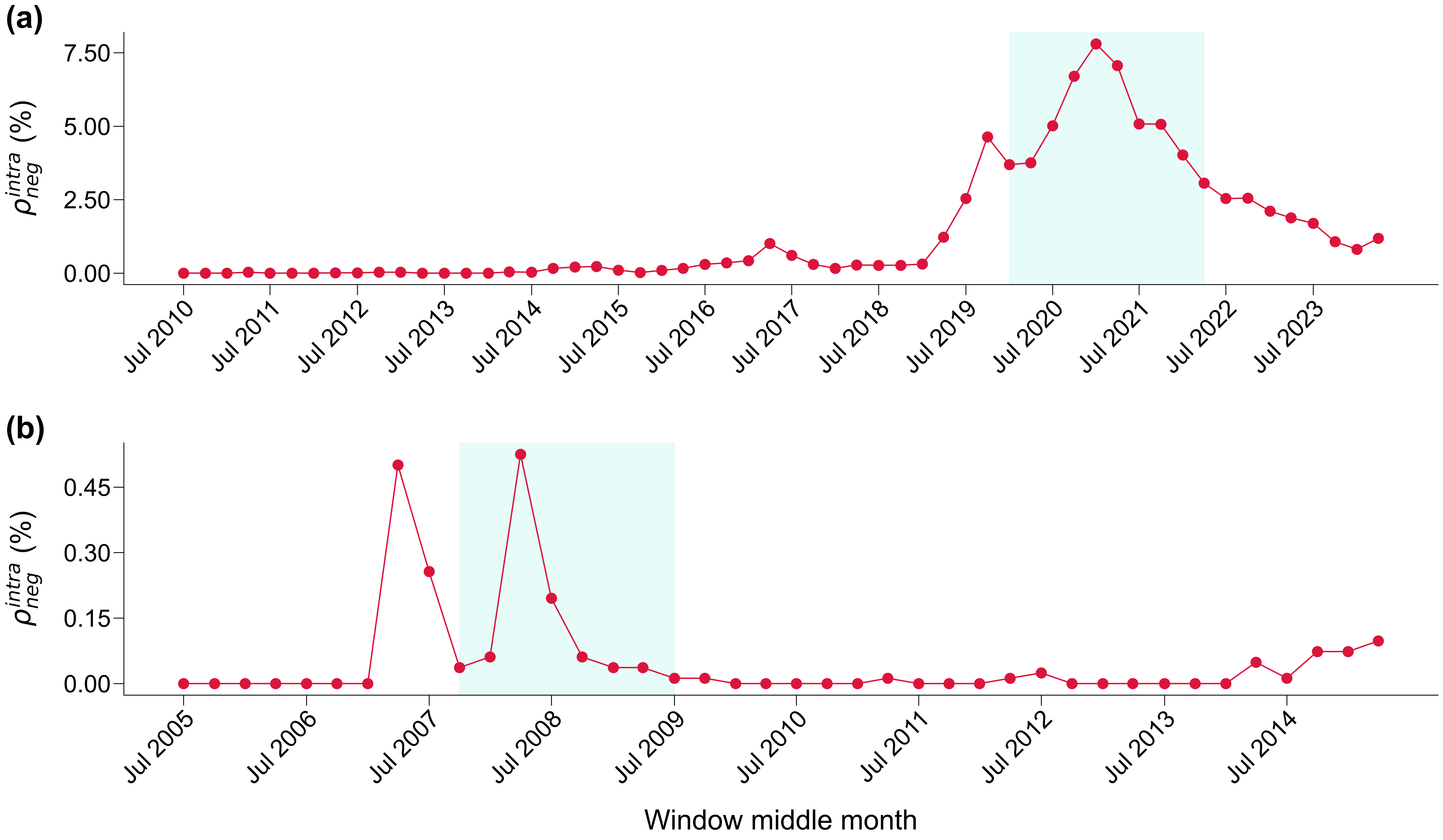}
    \caption{\textbf{Temporal evolution of the fraction of negative intra-sector edges.} (a) Results for the COVID-19 crisis pipeline. (b) Results for the 2008 Global Financial Crisis pipeline. The fraction is computed as the ratio of the total number of observed negative intra-sector edges within the sector to the total number of possible intra-sector edges. See formulation in SI Text S3.}
    \label{fig:rho_intra_neg_link_frac}
\end{figure}

One of the most striking discoveries in our dataset is the distinct structural difference between the 2020 COVID-19 crisis and the 2008 GFC. During the GFC, we observed a complete absence of intra-sector frustrated triads (see SI Figure S1), whereas they emerged uniquely during the COVID-19 crisis. This is confirmed by our analysis of intra-sector negative edge density (see \autoref{fig:rho_intra_neg_link_frac}): Throughout the COVID-19 period, negative edges within sectors remained substantial, fluctuating between 2\% and 8\%. In contrast to the GFC period, when they were minimal, peaking at only 0.5\%. For statistical validation of the observed distinction in intra-sector network configuration between the two crises, we conducted an exact permutation test on the peak magnitudes of intra-sector negative edge densities $(\rho_{neg}^{intra})$ during the core phases of the 2008 GFC and the COVID-19 crisis. Particularly, the first seven peak values of $\rho_{neg}^{intra}$ corresponding to each crisis period, yielding two samples of equal size ($n=7$). Under the null hypothesis that both crises arise from the same underlying network topology. We then compute the empirical difference in mean densities and compare it against the exact permutation distribution ($K = \binom{14}{7}=3,432$ combinations). A one-sided test is performed to evaluate whether the mean intra-sector negative edge density during the COVID-19 crisis is greater than that during the 2008 GFC. The test yields a highly significant $p$-value of $2.9\times10^{-4}$, providing strong evidence against the null hypothesis and confirming that both crises possess distinct intra-sector topologies. This change in triadic topology may likely be associated with the nature of the crisis: the 2008 Global Financial Crisis was an endogenous shock originating from inequalities across different parts of the economy. 
Conversely, the COVID-19 crisis was characterized by its exogenous nature, in which government-mandated lockdown disrupted companies within the same sector in very different ways~\cite{agaton2024market}. 

The multiscale analysis reveals a clear hierarchy in the organization of structural frustration within the S\&P 500. At the mesoscopic level, sectors remain internally coherent and largely balanced across most sectors. At the intermediate scale, however, interactions between sectors become increasingly heterogeneous and frustrated. The frequency analysis of the negative inter-sector edge and the associated negative partial correlation, representing the direct asset-level divergence while controlling for the S\&P 500 index, reveals the underlying mechanism leading to the loss of structural balance in the representative sector pairs shown in \autoref{fig:p_inter}. As established, we observe a pronounced decline in $P^{inter}$ during 2016-2017 due to a negative relative correlation between digital platforms (e.g., GOOGL) and Retail REITs (e.g., KIM, SPG). This opposing movement is most plausibly associated with the 2017 retail apocalypse, a period characterized by the closure of high-profile retailers and the growth of e-commerce. A second major decline in the Real Estate--Communication Services sector occurred during 2021-2022, which may reflect the post-pandemic recovery phase, as global venues reopened, entertainment stocks (e.g., LYV) experienced rising demand, against self-storage REITs (e.g., EXR, PSA), which previously served stable returns. Finally, we observe negative partial correlation between Oil \& Gas E\&P stocks (e.g., COP, EOG) and Drug Manufacturing stocks (e.g., JNJ), likely associated with the onset of the Russia--Ukraine conflict, with the Healthcare sector experiencing adverse effects while the Energy sector remained comparatively resilient.

The inconsistencies in inter-sectoral relationships accumulate to produce the fluctuations observed in global polarization. Our observations on global polarization align with the recent findings reported by Mart{\'\i}nez-Ramos \emph{et al.}~\cite{martinez2024covid}, who utilized k-means clustering on short 20-day epochs to identify a previously unseen "COVID anomaly state" in the correlation matrix, which appears around June 2020 and tapered off by February 2022. Correspondingly, we observe a sustained collapse and subsequent recovery in global polarization over this duration. Furthermore, while the employed clustering methodology categorizes market behavior into strictly discrete states, our decomposed $P_G$ metric tracks the continuous evolution of the network. The temporal overlap between their discrete anomalous state and the continuous decline in $P_G$ confirms that both approaches capture the same underlying structural transition despite operating at different levels of granularity.

The utilized measure of polarization not only assesses topological balance but moves beyond pure structural assessments by embodying the energy landscape framework~\cite{marvel2009energy}. In this context, we find that interactions within the same sector consistently reach a low-energy state, suggesting a tendency toward the global minimum of the sectoral energy landscape. Conversely, global-scale analysis reveals that while the system maintains a stable, low-energy organization during periods of market stability, it becomes trapped in local minima or jammed states during systemic crisis. This mirrors recent empirical findings in functional brain networks~\cite{saberi2024brain, saberi2025empirical}, highlighting the universality of these phase transitions across diverse complex systems.

Our final analysis provides evidence that the evolution of global polarization is closely associated with macroeconomic variables measuring the aggregate economy-wide trends. Specifically, the regression model shows that 68.2\% of the variance in global polarization is explained by the combined effects of statistics derived from the global supply chain index (GSCPI) and the consumer price index (CPI). The GSCPI is a composite index used to capture supply chain disruptions arising from the physical transport of goods, based on a range of indicators such as delivery time, changes in shipping rates, and air freight costs whose elevated levels indicate heightened supply disruptions~\cite{benigno2022gscpi}. Moreover, CPI is a price index that measures the percentage change in the price of a basket of goods that a consumer pays for and serves as a key indicator of inflation, while its standard deviation represents the magnitude of inflation uncertainty over the period considered~\cite{nguyen2023consumer}. We observed statistically significant negative coefficients for explanatory variables derived from the selected macroeconomic variables with respect to global polarization, suggesting that the structural imbalance in the market during the COVID-19 pandemic was strongly associated with physical bottlenecks and economic uncertainty.

\section{Conclusion}

In conclusion, we have demonstrated that the loss of global structural balance during periods characterized by systemic risk, in the S\&P 500 network representing interaction within the components of a complex financial system is fundamentally driven by conflicts between sectors, rather than instability within individual sectors. To quantify the respective contributions of intra- and inter-sector imbalance to the global network structure, we extract the mesoscopic group structure from the cross-correlation network and employ an order parameter, inspired by statistical mechanics, to characterize the critical transitions experienced by the system. Our balance-theoretic analysis of the resulting group structure reveals that many sectors within the S\&P 500 exhibit a cooperative network organization regardless of prevailing economic conditions,  whereas interactions between sectors are characterized by an incompatible network organization. Furthermore, our analytical framework yields qualitatively consistent results for the 2008 Great Financial Crisis, demonstrating its ability to identify the underlying nature of systemic economic disruptions. Benchmarking the empirical network against the proposed null models confirms that the observed structural balance, quantified through closed triadic motifs, is statistically significant and cannot be attributed to random network organization. In addition, regression analysis of the relationship between the employed order parameter and the selected macroeconomic variables yields robust regression estimates. Overall, this framework deepens our understanding of the long-term collective dynamics in the complex financial system and offers a principled approach to understanding the mechanisms by which inter-sector conflict accumulates and propagates, ultimately contributing to large-scale cascading failures during periods of systemic risk. Moreover, the method developed in this work is general and can be readily applied to other financial markets, such as foreign exchange~\cite{chakraborty2018deviations, chakraborty2020uncovering} and cryptocurrency markets~\cite{shukla2026time}.

\section*{Acknowledgements}
We gratefully acknowledge the valuable suggestions of Sitabhra Sinha and Tapan C. Adhyapak during the early stages of this work. We also thank Anindya S. Chakrabarti for his constructive and insightful comments on an earlier version of the manuscript.



\clearpage
\section*{Supplementary Information}

\beginsupplement


\section{Robustness check: Structural balance during the 2008 global financial crisis}
\label{sec:robustness_check_2008gfc}
\subsection{Dataset collection and network construction overview}

To validate the complete analytical pipeline employed for the investigation of loss of structural balance during the COVID-19 crisis (2010-2024), we further evaluate the developed framework on an independent dataset covering a comparable period characterized by both stable and unstable market conditions, namely the 2008 Global Financial Crisis (GFC). For this cross-validation, we utilize the daily adjusted closing prices of the companies listed on the S\&P 500 obtained from \href{https://finance.yahoo.com/}{Yahoo Finance}~\cite{yahoofinance_covid_pipeline} from all available tickers corresponding to S\&P 500 constituents, spanning January 1, 2005 to December 31, 2015. This results in an initial dataset of 465 stocks over a ten-year horizon. To ensure data quality, only stocks with less than 5\% missing observations are retained, resulting in a final dataset of 392 stocks. Additionally, we map all the companies in this time frame to the modern 11 GICS sector~\cite{msci2023gics} adopted in the COVID-19 analysis, thereby ensuring consistency between the sector-level analyses of the two study periods. Further, we segment the complete timeline into 40 overlapping windows using the same configuration adopted for the COVID-19 analysis. The window length $(T)$ is 252 days and has a 63-day overlap between consecutive windows. Missing values within each window were forward-filled with the last available closing price, treating them as non-trading days. For missing values occurring at the beginning of a time series, where no previous observations are available, backward filling is applied. Subsequently, the price series is converted into logarithmic returns (\autoref{eq:log_ret} in the main text) and the resulting log-return series are normalized according to \autoref{eq:norm_logret} in the main text. For each window, an interaction network is constructed by computing equal-time cross correlation matrix whose elements are defined by \autoref{eq:cross_corr} in the main text. Group correlation matrices are reconstructed using RMT-based filtering procedure, followed by the same thresholding approach described in \autoref{eq:thresholding} in the main text. The resulting signed interaction networks are analyzed using the same pipeline employed in the COVID-19 study to quantify the loss of structural balance both across and within market sectors through the polarization metrics proposed \autoref{subsec:sbt_quantified} in the main text. 

\subsection{No evidence of intra-sector frustrated triads during the 2008 GFC.}

\begin{figure}[!htbp]
    \centering
    \includegraphics[width=\textwidth]{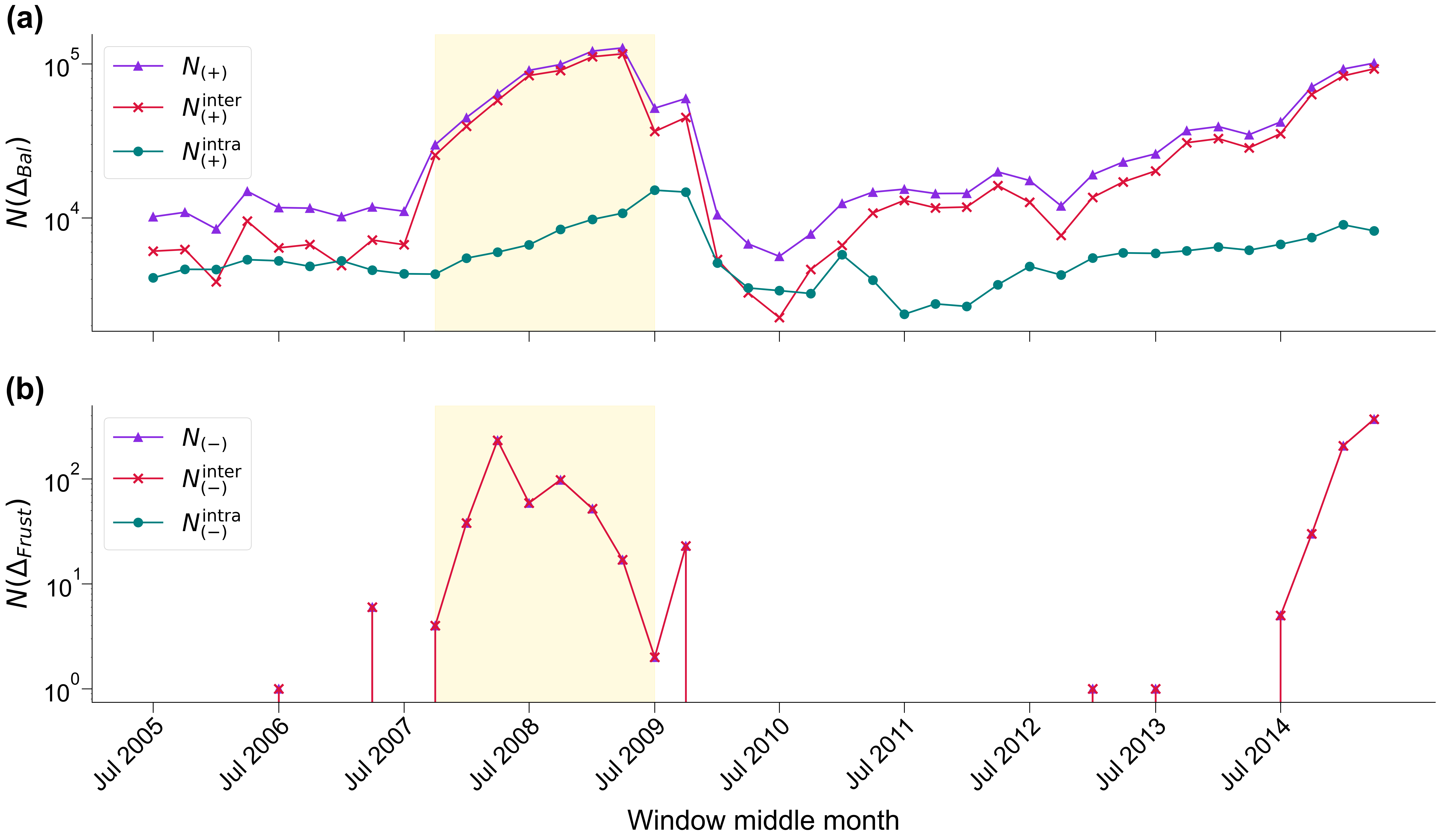}
    \caption{\textbf{Temporal evolution of balanced $N(\Delta_{Bal})$ and frustrated $N(\Delta_{Frust})$ triads, resolved into intra-sector and inter-sector configurations.}  (a) shows the total number of balanced triads $(N_{+})$, along with the numbers of inter-sector $(N_{+}^{{inter}})$ and intra-sector $(N_{+}^{{intra}})$ balanced triads. (b) shows the corresponding quantities for frustrated triads, namely the total number $(N_{-})$ and its inter-sector $(N_{-}^{{inter}})$ and intra-sector $(N_{-}^{{intra}})$ components. The shaded region denotes the 2008 GFC, spanning from window middle date Jan 03, 2008 to July 06, 2009 .}
    \label{fig:supp_bal_frust_sector_triads_evol_gfc}
\end{figure}

The temporal evolution of the number of balanced triads in the S\&P 500, shown in \autoref{fig:supp_bal_frust_sector_triads_evol_gfc}(a), demonstrates that balanced triads of all categories persist and outnumber frustrated triads throughout the observation period. In general, the number of inter-sector balanced triads ($N_{+}^{inter}$) exceeds that of intra-sector balanced triads ($N_{+}^{intra}$), except at during few epochs where this trend is reversed. In contrast, the temporal evolution of frustrated triads shown in \autoref{fig:supp_bal_frust_sector_triads_evol_gfc}(b) exhibits pronounced fluctuations only during the 2008 GFC, which is then followed by a complete absence before re-emerging after 2014. The number of frustrated triads observed during the GFC is consistent with the findings reported by Kuyyamudi \emph{et al.}~\cite{kuyyamudi2019emergence}, providing independent validation of the observed count of frustrated triads during this crisis period. Although the increase in inter-sector frustrated triads confirms the presence of a system-wide structural disruption during the GFC, no evidence of intra-sector frustrated triads is observed during this crisis. As discussed earlier, this is attributed to the substantially lower density of intra-sector negative edges observed during GFC compared to COVID-19 crisis (see \autoref{fig:rho_intra_neg_link_frac} in main text).

\subsection{Polarization findings on 2008 GFC timeline}

\begin{figure}[!htbp]
    \centering
    \includegraphics[width=\textwidth]{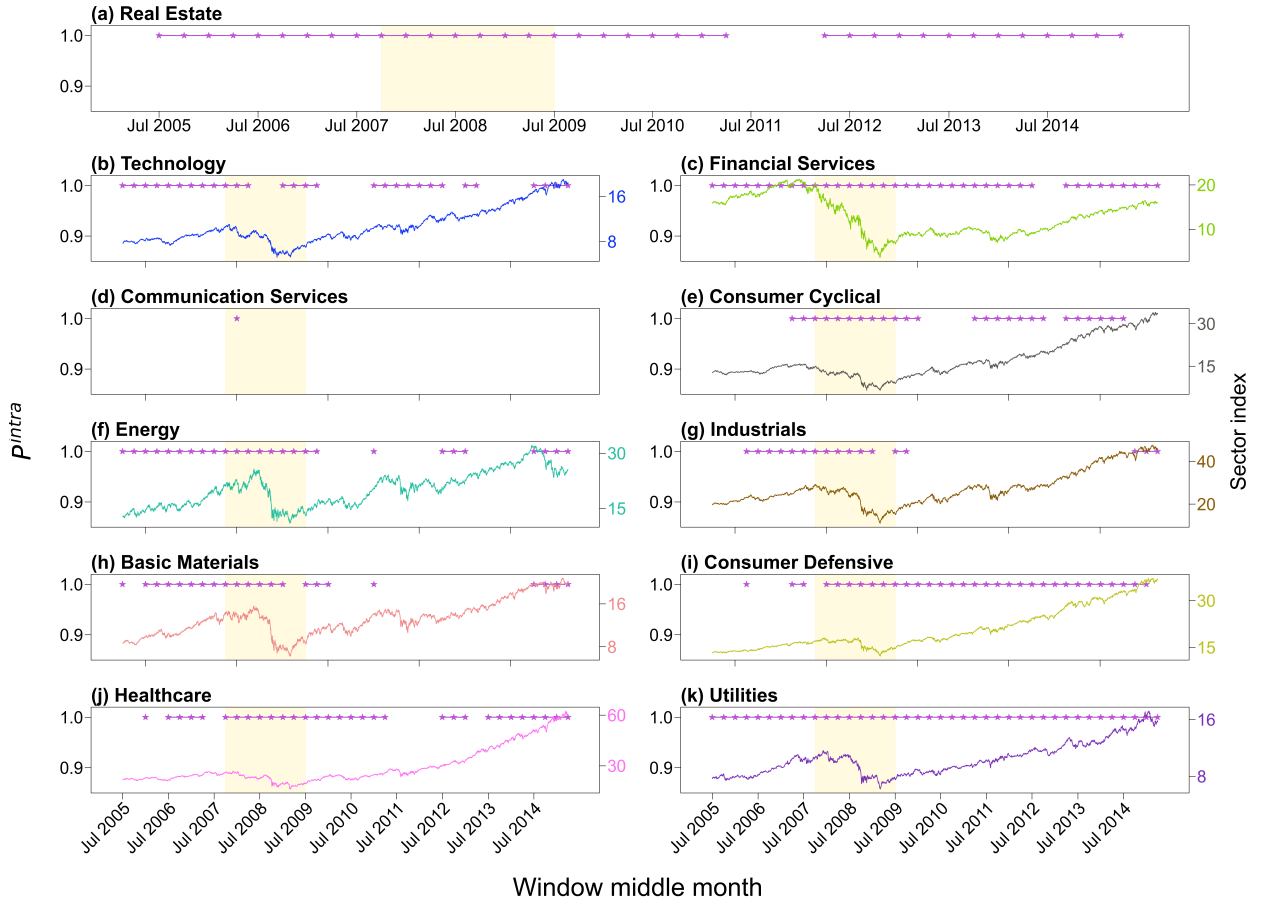}
    \caption{\textbf{Temporal evolution of intra-sector polarization across the 11 GICS sectors.} (a)–(k) present $P^{intra}$, for each of the 11 GICS sectors in the S\&P 500 over the study period. The corresponding sector index is shown on the secondary axis in each subplot. The shaded area indicates the 2008 GFC region. Missing values of $P^{intra}$ correspond to periods in which no triads were formed within the sector, rendering the polarization measure undefined. The sector indices for Real Estate and Communication Services are omitted because these sectors were not classified as independent sectors prior to 2015.}
    \label{fig:supp_p_intra}
\end{figure}

At the intra-sector level, the polarization of the sector-specific interaction networks, classified according to the modern 11 GICS sectors, is shown in \autoref{fig:supp_p_intra}. We observe a state of complete structural balance within the sector -specific networks, irrespective of prevailing market conditions. This observation is corroborated by the complete absence of intra-sector frustrated triads throughout the study period. These finding also aligns with the strongly polarized states obtained from the COVID-19 analysis, where the majority of sector-specific networks maintained a high degree of polarization. From the perspective of structural balance theory, this observation suggests internally consistent interaction among the constituents of the same group. 

\begin{figure}[!htbp]
    \centering
    \includegraphics[width=\textwidth]{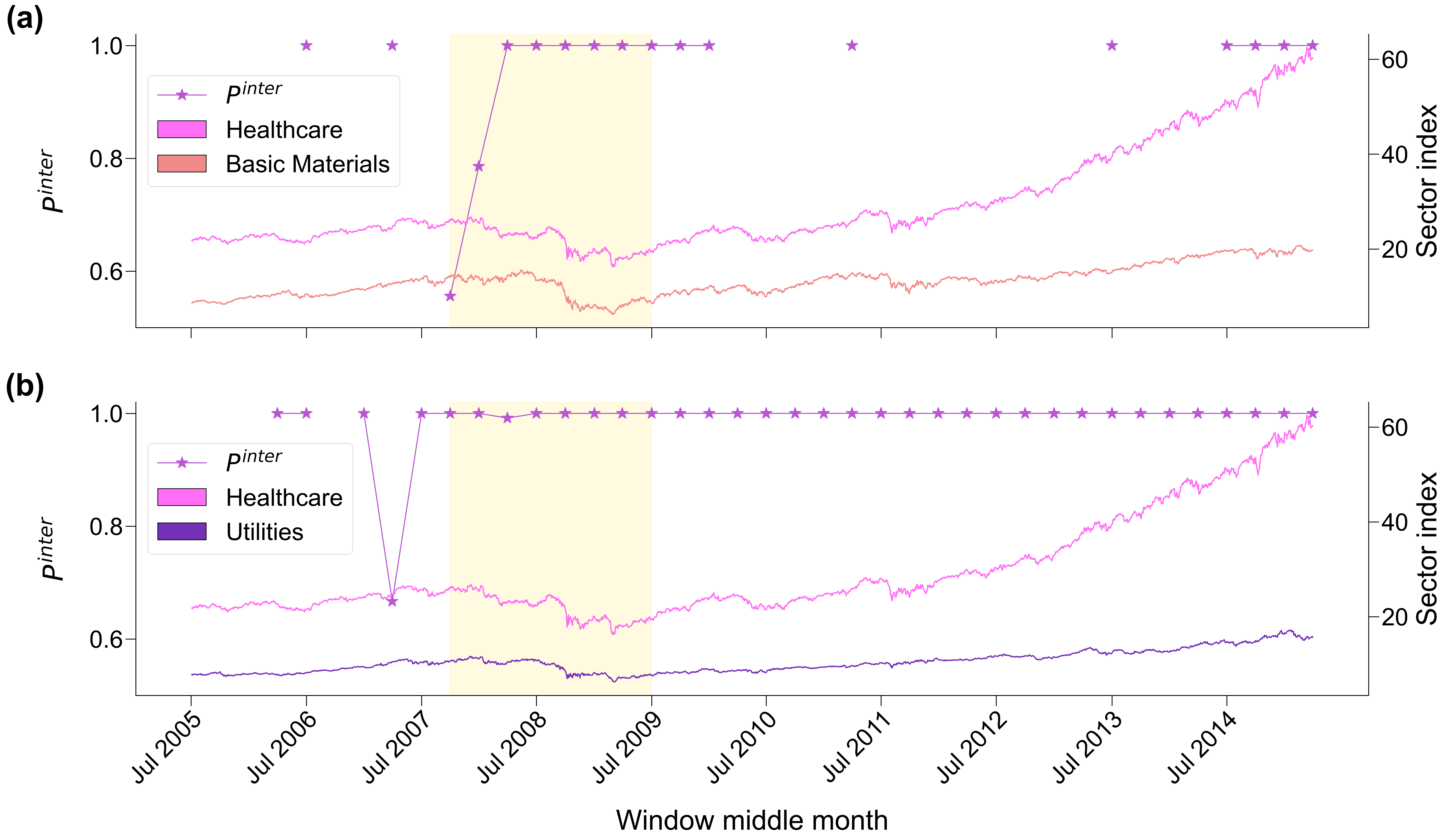}
    \caption{\textbf{Temporal evolution of inter-sector polarization $(P^{inter})$ for the two least-polarized sector pairs.} Sector indices are shown on the secondary axis, and the shaded area indicates the 2008 GFC region. $P^{inter}$ is undefined during periods in which no inter-sector triads are formed, resulting in gaps in the curves.}
    \label{fig:supp_p_inter}
\end{figure}

\autoref{fig:supp_p_inter}, illustrates the inter-sector polarization for the two representative sector pairs (selected from the 55 possible combinations) exhibiting the lowest levels of polarization over the study period. Among these, the minimum polarization is observed for the Healthcare and Basic Materials pair. In contrast to the intra-sector networks, the loss in structural balance becomes evident when interaction between sectors is considered, indicating that the structural imbalance in the market resides at the boundaries between fundamentally opposed sectors. Furthermore,  the inter-sector polarization during the 2008 GFC remains comparatively stable relative to the several declines in polarization observed during the COVID-19 period. This suggests that although both crises disrupted the inter-sector interaction structure, the COVID-19 crisis induced a greater departure from structural balance than the 2008 GFC. 

\begin{figure}[!htbp]
    \centering
    \includegraphics[width=\textwidth]{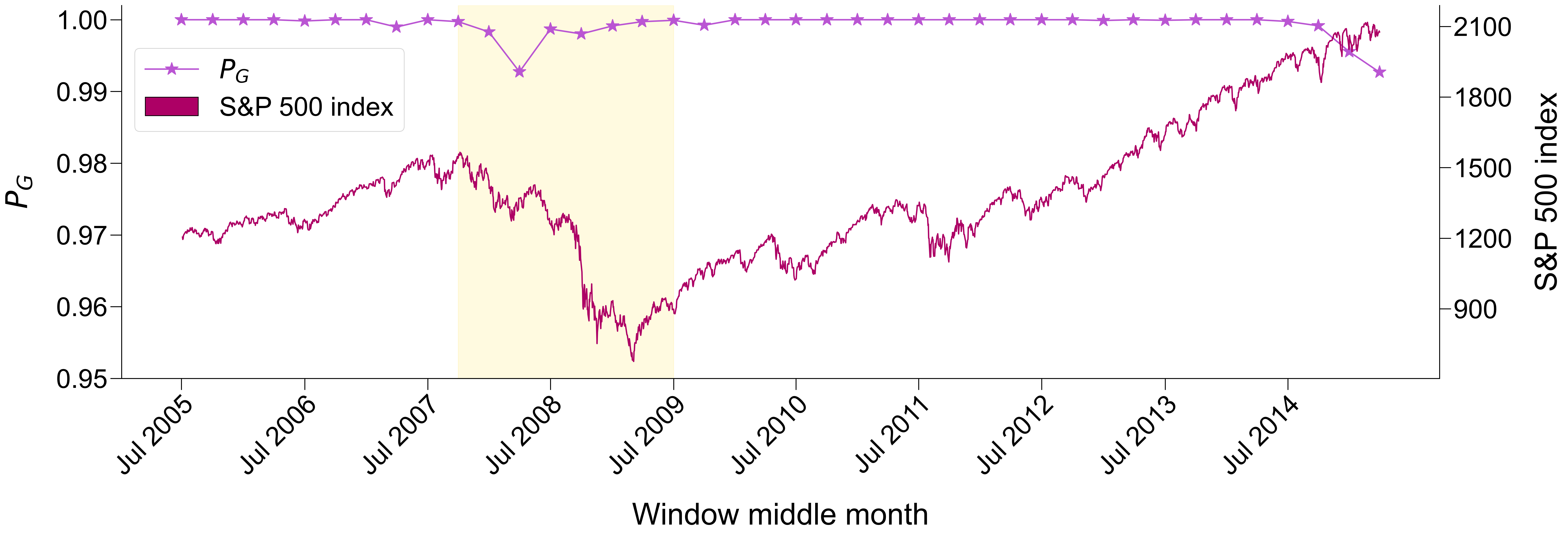}
    \caption{\textbf{Temporal evolution of global polarization.} The S\&P 500 index is shown on the secondary y-axis; the shaded area denotes the 2008 GFC region.}
    \label{fig:supp_pg}
\end{figure}

At the global scale, as shown in \autoref{fig:supp_pg}, the decline in polarization observed at the onset of the 2008 GFC indicates the loss of structural balance across the interaction network. As the crisis subsides, the network gradually returns to a more balanced state, as indicated by the recovery in polarization levels. By the end of the crisis period, the network returns to a highly polarized, structurally balanced state, which is subsequently maintained throughout the remainder of the study period. Overall, the polarization dynamics across the global, inter-sector, and intra-sector scales are fully consistent with those observed during the COVID-19 period. In both cases, major crisis events are characterized by a reduction in polarization, whereas prolonged periods of market stability correspond to a fully polarized network ($P_G = 1$), indicative of complete structural balance. The consistency of these observations across two independent datasets robustly validates our proposed analytical framework. 

\subsection{Null models}

\begin{figure}[!htbp]
    \centering
    \includegraphics[width=\textwidth]{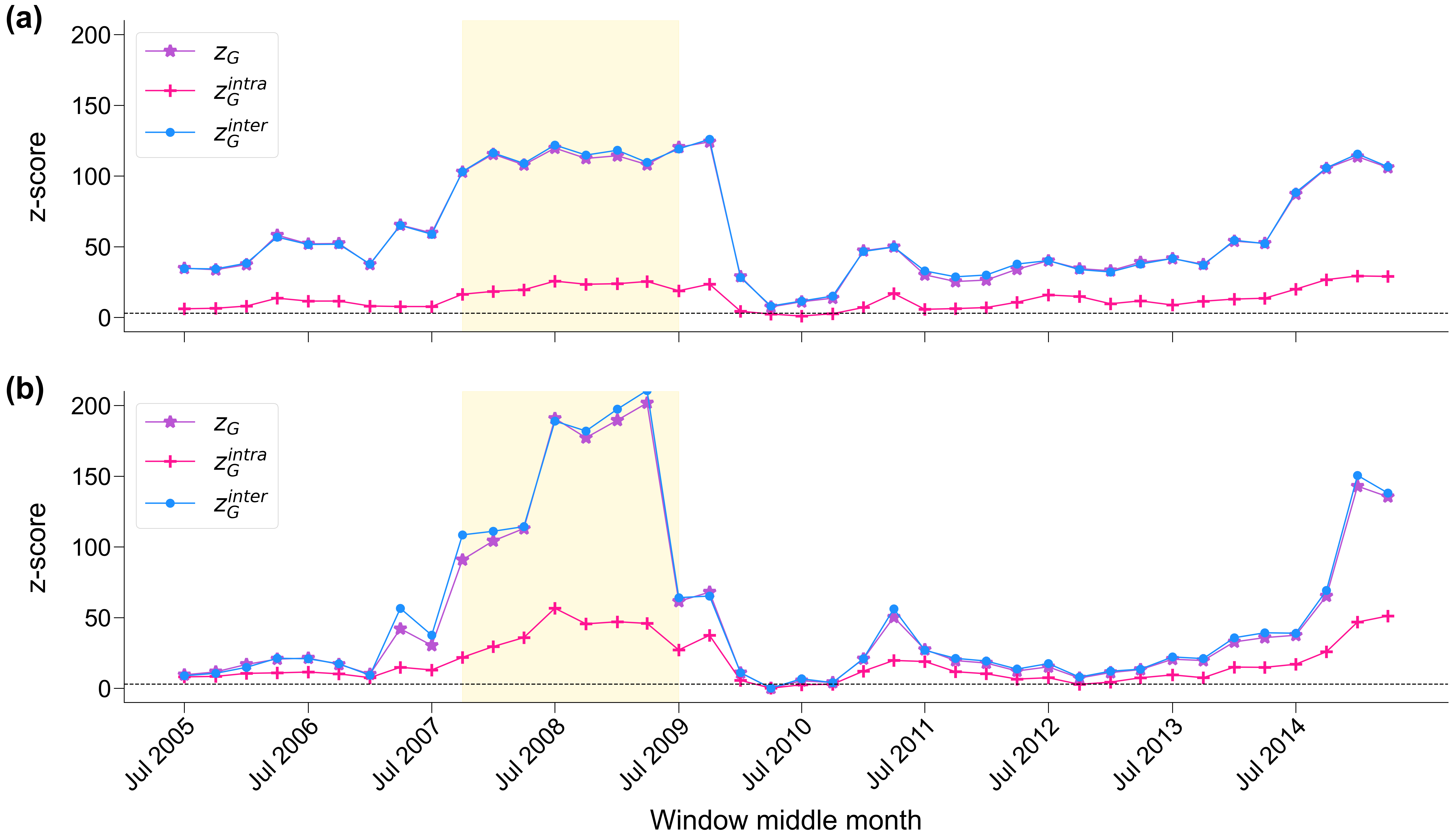}
    \caption{\textbf{Temporal evolution of the z-scores for the global ($P_G$), intra-sector ($P_{G}^{intra}$), and inter-sector ($P_{G}^{inter}$) polarization metrics} under (a) the sign-rewiring null model and (b) the signed topology-preserving (STP) maximum-entropy null model. The horizontal dotted line at $z=3$ marks the threshold for statistical significance, and the shaded region indicates the 2008 GFC period.}
    \label{fig:supp_null_models}
\end{figure}

To assess the statistical significance of the 2008 GFC structural balance, we benchmarked the empirical polarization metrics against both the sign-rewire~\cite{maslov2002specificity} (see \autoref{fig:supp_null_models}(a) and the signed topology preserving null model~\cite{hao2024proper} (see \autoref{fig:supp_null_models}(b)). During the crisis period, the z-scores show massive spikes, indicating a highly non-random structure. We observed a few cases in each null model exhibiting a drop below the significance threshold primarily during the post-crisis phase, indicating those specific windows where the empirical network remained statistically indistinguishable from the null model. For the global intra-sector polarization ($P_{G}^{intra}$), drops below $z=3$ occurred under both null models. Nevertheless, the global polarization metrics in general remained statistically significant across majority of the periods under study.

\section{Frequency of inter-sector negative edges during the temporal evolution of $P^{inter}$}
\label{sec:partial_corr_tables}
To investigate the major declines in the inter-sector polarization measure $(P^{inter})$ observed in \autoref{fig:p_inter} (in \autoref{sec:all_polarization_findings} of the main manuscript), we analyzed the frequency with which individual stock pairs formed the negative edges in the imbalanced triads during the periods corresponding to these declines. Further, we estimated the partial correlation coefficient for the identified stock pairs to quantify the direct asset-level divergence while controlling for common market movements~\cite{baba2004partial, kenett2010dominating}. The partial correlation coefficient quantifies the direct linear relationship between the returns of the two stocks while controlling for the returns of the S\&P 500 index. This ensures the removal of the linear contribution of the market index and thus captures the residual pairwise relationship between the two stocks. The partial correlation $\rho_{i,j:SP}$ between two log return time series $R_i$ and $R_j$ with respect to the S\&P 500 index log returns $R_{SP}$ is defined as,

\begin{equation}
\rho_{i,j:SP}
=
\frac{C_{ij}-C_{i,SP}C_{j,SP}}{\sqrt{\left(1-C_{i,SP}^2\right)\left(1-C_{j,SP}^2\right)}}.
\end{equation}

Here, $C_{ij}$ denotes the Pearson correlation between the log returns of stocks $i$ and $j$, while $C_{i,SP}$ and $C_{j,SP}$ denote the Pearson correlations of stocks $i$ and $j$ with the S\&P 500 index log returns, respectively. 

\autoref{tab:partial_corr_tb1}--\autoref{tab:partial_corr_tb3} summarize the details of the inter-sector negative edges observed during periods of pronounced decline in inter-sector polarization. Each table reports the top five stock pairs forming inter-sector negative edges, the industries to which the constituent stocks belong, the frequency with which each edge appears in the identified imbalanced triads, and the partial correlation coefficient.

\begin{table}[!htbp]
\caption{Top five highest-frequency inter-sector negative edges between the Communication Services and Real Estate sectors during the Oct 2016--Oct 2017 window when $P^{inter}$ decreases to $0.69$. The combined sector subgraph period contains 136 $(+++)$, 29 $(++-)$, 27 $(+--)$, and no $(---)$ triads. In each ticker pair, the first ticker (GOOG/GOOGL) belongs to the Communication Services sector, while the second ticker (KIM, REG, and SPG) belongs to the Real Estate sector.}
\centering          
\footnotesize         
\renewcommand{\arraystretch}{2} 
\setlength{\tabcolsep}{7pt} 
\begin{tabular}{
        l
        >{\centering\arraybackslash}p{4.8cm} 
        c
        S[table-format=-1.4, round-mode=places, round-precision=3]
}
\toprule 
{Ticker pair} & 
{Industry} &
\makecell{Edge frequency} & 
\multicolumn{1}{c}{\makecell{Partial correlation}} \\
\midrule 
(GOOGL, KIM) & (Internet Content \& Information, REIT Retail) & 5  & -0.272569721 \\
(GOOGL, REG) & (Internet Content \& Information, REIT Retail) & 5  & -0.215504793 \\
(GOOGL, SPG) & (Internet Content \& Information, REIT Retail) & 5  & -0.179296061 \\
(GOOG, KIM)  & (Internet Content \& Information, REIT Retail) & 3  & -0.294164981 \\
(GOOG, REG)  & (Internet Content \& Information, REIT Retail) & 3  & -0.248603954 \\
\bottomrule 
\end{tabular}
\label{tab:partial_corr_tb1}
\end{table}

\begin{table}[H]
    \caption{Top five highest-frequency inter-sector negative edges between the Communication Services and Real Estate sectors during Jan 2021--Jan 2022 when $P^{inter}$ decreases to $0.86$. The combined sector subgraph period contains 269 $(+++)$, 31 $(++-)$, 172 $(+--)$, and no $(---)$ triads. In each ticker pair, the first ticker (LYV, MTCH, DIS) belongs to the Communication Services sector (Internet Content \& Information), while the second ticker (EXR, PSA, CCI, DLR) belongs to the Real Estate sector.}
    \centering
    \footnotesize
    \renewcommand{\arraystretch}{2}
    \setlength{\tabcolsep}{7pt}
    \begin{tabular}
        {
        l
        >{\centering\arraybackslash}p{4.8cm}
        c
        S[table-format=-1.4, round-mode=places, round-precision=3]
        }
    \toprule
        {Ticker pair} & 
        {Industry} &
        {Edge frequency} & 
        {Partial correlation} \\
    \midrule
        (LYV, EXR)  & (Entertainment, REIT Industrial) & 5  & -0.10690163 \\
        (LYV, PSA)  & (Entertainment, REIT Industrial) & 4  & -0.205450776 \\
        (LYV, CCI)  & (Entertainment, REIT Specialty) & 2  & -0.159078934 \\
        (MTCH, EXR) & (Internet Content \& Information, REIT Industrial) & 2  & -0.179521017 \\
        (DIS, DLR)  & (Entertainment, REIT Specialty) & 2  & -0.168857256 \\
    \bottomrule
    \end{tabular}
    \label{tab:partial_corr_tb2}
\end{table}

\begin{table}[!htbp]
    \caption{Top five highest-frequency inter-sector negative edges between Energy and Healthcare sectors when $P^{inter}$ decreases to $0.77$. The combined sector subgraph period contains 483 $(+++)$, 153 $(++-)$, 796 $(+--)$, and 7 $(---)$ triads. In each ticker pair, the first ticker (COP, EOG, XOM, CVX) belongs to the Energy sector, the second ticker (JNJ, GILD) belongs to the Healthcare sector.}
    \centering
    \footnotesize
    \renewcommand{\arraystretch}{2}
    \setlength{\tabcolsep}{7pt}
    \begin{tabular}
        {
        l
        >{\centering\arraybackslash}p{4.8cm}
        c
        S[table-format=-1.4, round-mode=places, round-precision=3]
        }
    \toprule
        {Ticker pair} & 
        {Industry} &
        {Edge frequency} & 
        {Partial correlation} \\
    \midrule
        (COP, JNJ)   & (Oil \& Gas E\&P, Drug Manufacturers - General) & 7  & -0.166592886 \\
        (EOG, JNJ)   & (Oil \& Gas E\&P, Drug Manufacturers - General) & 7  & -0.176464433 \\
        (XOM, JNJ)   & (Oil \& Gas Integrated, Drug Manufacturers - General) & 7  & -0.15128648 \\
        (COP, GILD)  & (Oil \& Gas E\&P, Drug Manufacturers - General) & 6  & -0.093738254 \\
        (CVX, GILD)  & (Oil \& Gas Integrated, Drug Manufacturers - General) & 6  & -0.090656343 \\
    \bottomrule
    \end{tabular}
    \label{tab:partial_corr_tb3}
\end{table}

\FloatBarrier
\section{Density of intra-sector negative edges.}
\label{sec:supp_rho_neg_intra_sec_edges}
To quantify the proportion of negative edges within the market sectors. We define the density of intra-sector negative edges as followed. For a given sector $p$, the maximum possible number of intra-sector edges is $E_p = N_p(N_p-1)/2$, where $N_p$ denotes the number of constituent nodes in that sector. Summing over all sectors provides the total number of possible intra-sector edges $E_{intra}$. The density of intra-sector negative edges $\rho_{neg}^{intra}$ is computed as the ratio of total number of observed intra-sector negative edges within sector to the total number of possible intra-sector edges.

\begin{equation}
\rho_{{neg}}^{intra}
=
\frac{1}{E_{intra}}
\sum_{p}
\sum_{i < j \in p}
\begin{cases}
1, & \text{if } A_{ij} < 0,\\
0, & \text{otherwise}.
\end{cases}
\end{equation}
where $A_{ij}$ denotes the $(i,j)$ entry of the signed adjacency matrix.


\end{document}